\documentclass[usenatbib]{mn2e}
\usepackage{graphicx}
\usepackage{txfonts}
\usepackage[english]{babel}
\usepackage{url}
\bibpunct{(}{)}{;}{a}{}{,}
\usepackage{bm}

\begin{document}
\title[Search for strong lenses in HST archive]{A seven square degrees survey for galaxy-scale gravitational lenses with the HST imaging archive \thanks{Based on observations made with the NASA/ESA Hubble Space Telescope, obtained from the data archive at the Space Telescope Science Institute. STScI is operated by the Association of Universities for Research in Astronomy, Inc. under NASA contract NAS 5-26555.}}
\author[R.S. Pawase et al.]{R.S. Pawase$^{1}$, F. Courbin$^{1}$, C. Faure$^{1}$, R. Kokotanekova$^{1}$ and G. Meylan$^{1}$\\
$^{1}$Laboratoire d'Astrophysique, Ecole Polytechnique F\'ed\'erale de Lausanne (EPFL), Observatoire de Sauverny, 1290 Versoix, Switzerland}

\date{}
\maketitle


\begin{abstract}
We  present the results of a visual search for galaxy-scale gravitational lenses in $\sim$7\degr$^2$ of  Hubble Space Telescope (HST) images. The dataset comprises the whole imaging data ever taken with the Advanced Camera for Surveys (ACS) in the filter F814W (I-band) up to August 31$^{\rm st}$, 2011, i.e. 6.03\degr$^2$  excluding the field of the  {\it Cosmic Evolution Survey} (COSMOS) which has been the subject of a separate visual search. In addition,  we have searched  for lenses in the whole  Wide Field Camera 3 (WFC3)  near-IR imaging dataset in all filters (1.01\degr$^2$)  up to the same date.  Our primary goal  is to provide a sample of lenses with a broad range of different morphologies and lens-source brightness contrast in order estimate a lower limit to the number of galaxy-scale strong lenses in the future Euclid survey in its VIS band. Our criteria to select lenses are purely morphological  as we do not use any colour or redshift information. The final candidate selection is  very conservative hence leading to a nearly pure but incomplete sample. 
We  find  49 new lens candidates:  40 in the ACS images and 9 in the WFC3 images. Out of these, 16 candidates are  secure lenses owing to their striking morphology, 21 more are very good candidates, and 12 more  have morphologies compatible with gravitational lensing but also compatible with other astrophysical objects such as ring and chain galaxies or mergers. Interestingly, some lens galaxies include low surface brightness galaxies, compact groups and mergers. The imaging dataset is  heterogeneous in depth and spans a broad range of galactic latitudes. It is therefore insensitive to cosmic variance and allows to estimate the number of galaxy-scale strong lenses on the sky  for a putative survey depth, which is the main result of the present work.  Because of the incompleteness of the sample, the estimated lensing rates  should be taken as lower limits. Using these, we anticipate that a 15 000\degr$^2$ space survey such as Euclid will find at least 60~000 galaxy-scale strong lenses down to a limiting AB magnitude of $I=24.5$ (10-$\sigma$) or $I=25.8$ (3-$\sigma$).
\end{abstract}


\begin{figure*}
 \includegraphics[width= 1.0\textwidth]{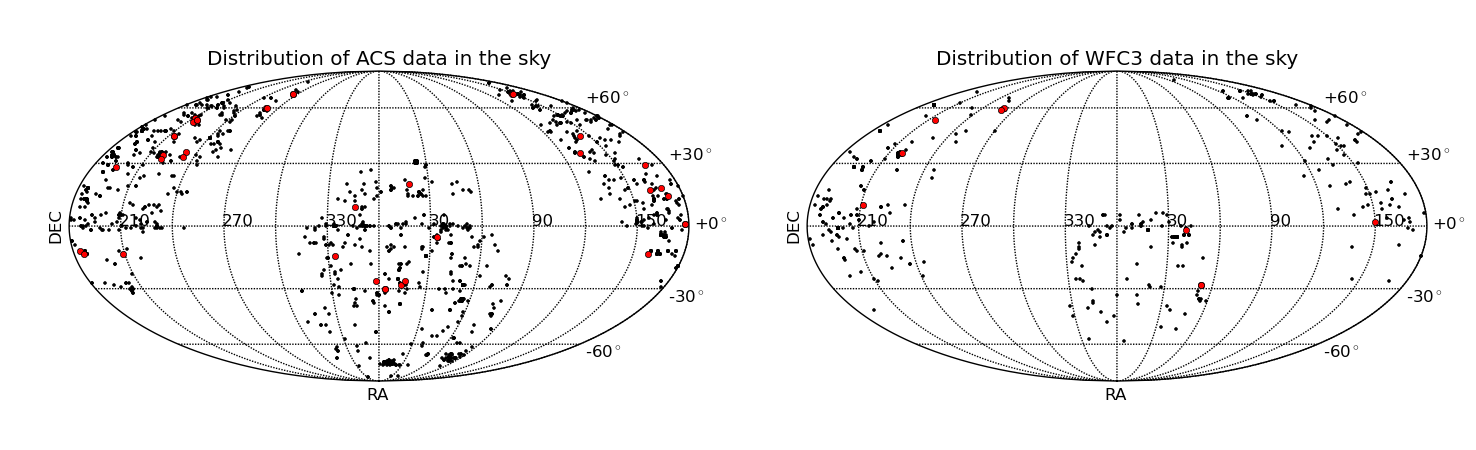}
 \caption{Distribution of the ACS  (left) and  WFC3 (right) images on the sky. The red dots indicate the position of the fields with a lens candidate. On the left panel, each point corresponds to a single ACS field of 11\arcmin$^2$. The total area covered by the ACS is 6.03\degr$^2$.  On the right panel, each point corresponds to a single WFC3 field of 4.65\arcmin$^2$. The total area covered by the WFC3 survey is 1.01\degr$^2$. 
} \label{fig:allsky}
\end{figure*}

\section{Introduction}

Gravitational lensing in its weak and strong regimes is currently one of the best tools to study dark matter and dark energy \citep[e.g.][]{Hu1999}. It is also the most reliable way to weigh precisely galaxies up to several effective radii  \citep[e.g.][]{Gavazzi2007}.  In combination with stellar dynamics, it allows us measure the dark and luminous mass profiles of galaxies \citep[e.g.][]{Barnabe2009} and to break the disc-halo degeneracy in spiral galaxies \citep[e.g.][]{Dutton2011}. 
 Thanks to large sample of strong lenses, statistical studies of galaxy mass properties and evolution with redshift are feasible \citep[e.g.][]{faure2011,faure2009,lagattuta2010,Auger2010, Koopmans2006}.  In some cases, when the radial extent of an Einstein ring is particularly large or when sources at multiple redshifts are lensed by the same object \citep[e.g.][]{Gavazzi2008}, the measurement of the mass slope in the lens can be of exquisite quality and makes it possible to constrain the cosmological parameters \citep[e.g.][]{Suyu2013, Tewes2012, Suyu2010} in a way fully competitive with other cosmological probes \citep[see][for a short comparison of the methods]{Suyu2012}. 

The increasing number of optical all-sky surveys either from the ground or from space, lends considerable hope to build such large samples, with tens or even hundreds of thousands of objects,  as will be illustrated in the present work. However, the automated techniques available to find the lenses from petabytes of imaging data are so far limited in their efficiency and tend to produce a large number of false positives that require significant post-processing and cleaning \citep[e.g.][]{Marshall2009}. 


Among the best automated robots to find lenses are {\tt Arcfinder} \citep{Seidel2007}, which was primarily developed to find large arcs behind clusters and groups, and the algorithm by \citet{Alard2006} used by \citet{Cabanac2007} and \citet{More2012} which was optimized to look for arcs produced by individual galaxies and groups in the CFHT Strong Lensing Legacy Survey (SL2S). While the latter techniques do not rely on any model for the mass distribution in lens galaxies, other automated robots consider any galaxy as a potential lens and predict where lensed images of a background source should be before trying to identify them on the real data \citep{Marshall2009}. Such robots will be mandatory to carry out lens searches in surveys of intractable size for a few astronomer , i.e. the all-sky surveys that will take place in the next decade. 


An alternative to robots will be the search  for strong  lenses by citizens in projects like  the Galaxy Zoo \citep{lintott2008} and Spacewarps (PI: P. Marshall). 
In the same spirit, visual inspections of HST archive images in search for strong lenses have been (successfully) attempted in the  past:  \citet{Ratnatunga1999} found 10 lenses in the HST medium deep survey,  followed by \citet{Fassnacht2006}, \citet{Moustakas2007} and \citet{Newton2009} who found more systems. So far, the largest search for strong lenses in HST images  is the one conducted in the 1.64\degr$^2$ field of view of the   {\it Cosmic Evolution Survey}  \citep[COSMOS;][]{scoville2007} by  \citet{Faure2008} and \citet{Jackson2008}. They found  in total 179  lens candidates among which 22 display multiple images of the source and/or  have both  lens and source redshifts to confirm their lens nature  \citep[see][]{faure2011}. 

In the present paper, we describe a search for strong lenses by visual inspection of all the HST images ever taken with the Wide Field Channel (WFC) of the Advanced Camera for Surveys (ACS) through the F814W filter (I-band), up to August $31^{\rm st}$, 2011. The total field of view explored is 6.03\degr$^2$ and {\it excludes} the COSMOS field, which has been used already in the past by two independent teams for the same purpose.  

Since future wide field surveys will include near-IR imaging, we also carry out the experiment using the data taken with the Wide Field Camera 3 (WFC3) in the near-IR channel and the F160W filter. As the WFC3 is a rather new instrument on the HST and since the field of view of the camera itself is limited, the total area we cover in the near-IR is smaller than with the ACS data, i.e., 1.01\degr$^2$. 

Our primary goal with this work is to estimate in an empirical way (i.e. not based on models but rather on data) the minimum number of galaxy-scale lenses that may be discovered in future whole-sky surveys as a function of depth. In addition, our new sample illustrates the broad diversity in image configurations, spatial scales and light contrasts between the lenses and sources. This should help designing automated robots by providing a broader variety of priors on the properties of lensing systems than previously known. Lens modeling, follow-up observations, and studying galaxy evolution are out of the scope of this work. Throughout the paper, all magnitudes are in the  AB magnitude system.

\begin{figure}
 \includegraphics[width= 0.5\textwidth]{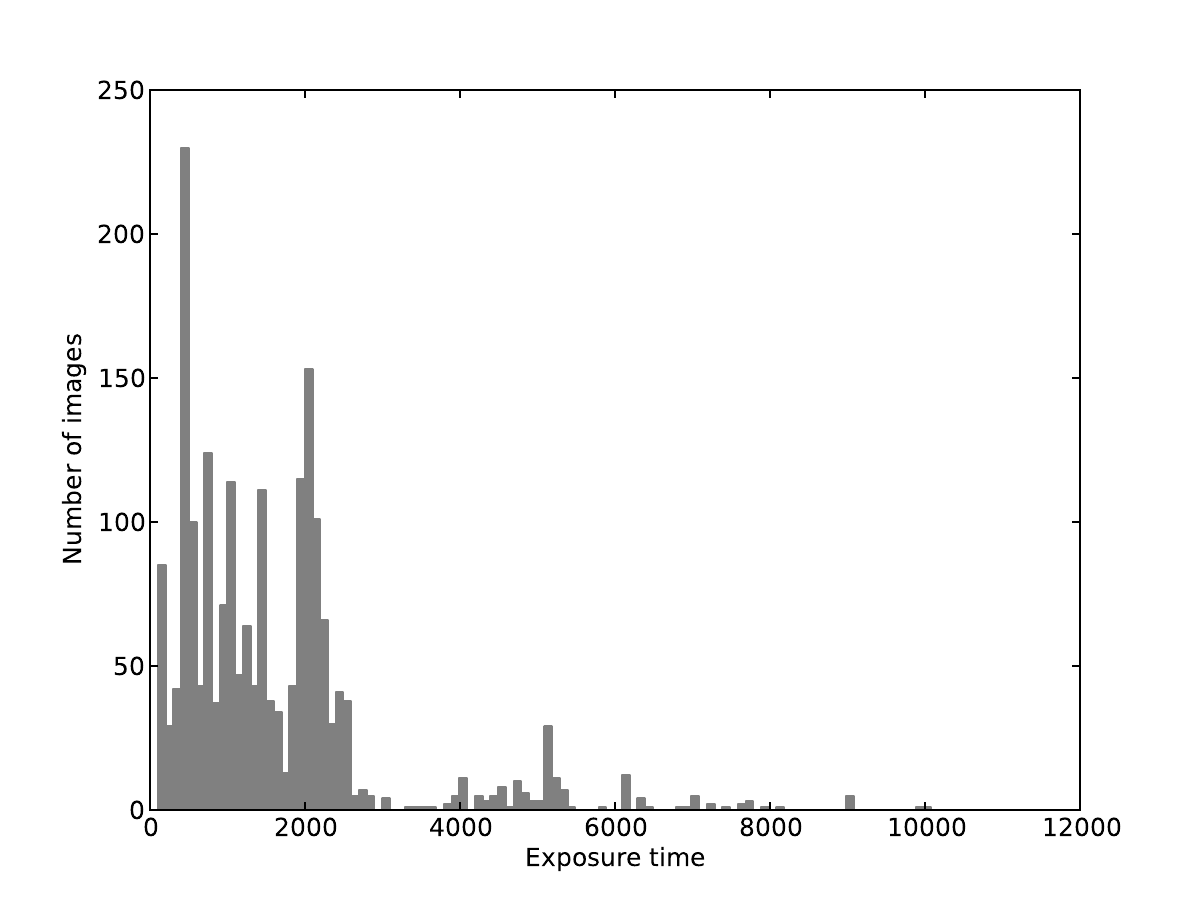}
 \caption{Distribution of the  ACS/F814W images exposure time.}
 \label{fig:histoexptimeacs}
\end{figure}

\section{Dataset and characteristics}

\begin{figure}
 \includegraphics[width= 0.52\textwidth]{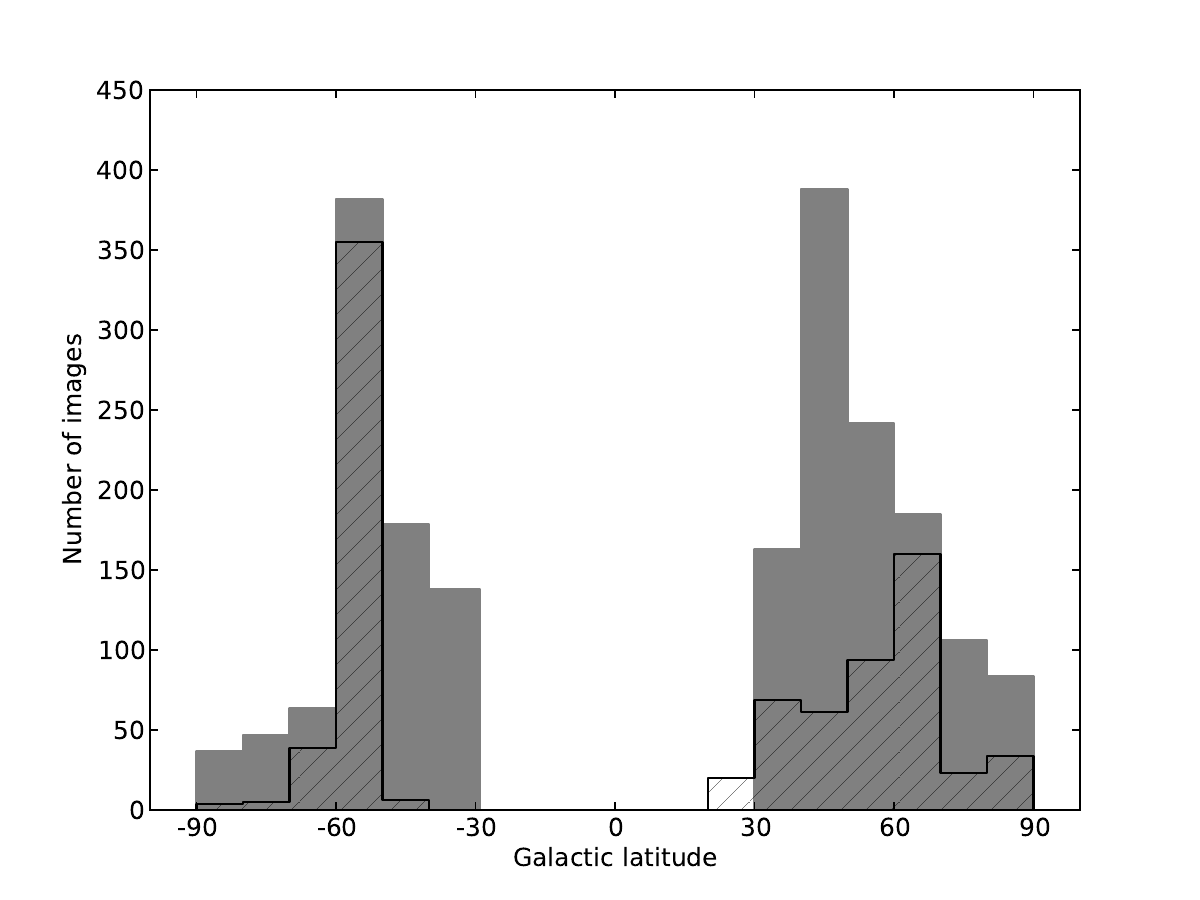}
 \caption{Distribution of the ACS (filled histogram) and WFC3 frames (dashed histogram) in galactic latitude.}
 \label{fig:histogalaclat}
\end{figure}

This work is all based on archival HST images from the Mikulski Archive for Space Telescopes (MAST) and its associated querying tool: the Hubble Data Archive (HDA)\footnote{http://archive.stsci.edu/hst/search.php}.

We  searched for strong lenses in all the  ACS imaging taken in the broadband I filter (F814W) and available in the archive on August $31^{\rm st}$, 2011. We limited our search to the range of galactic latitude where the Milky Way does not lead to severe contamination by foreground objects and where the Galactic absorption is minimized, i.e., $l=[+30^\circ;+90^\circ]$  and $l=[-90^\circ;-30^\circ]$. The final calibrated, geometrically-corrected, dither-combined images with a pixel scale 
0.05\arcsec /pixel were  used for inspection.  They cover a total of 6.03\degr$^2$ of the sky. 

We also carried out a near-IR search, using all the WFC3 image available in the archive on August $31^{\rm st}$, 2011.  Data were available in ten different filters (see Table~\ref{tab:wfc3filter}) with Galactic latitudes in the range $l=[+20^\circ;+90^\circ]$ and $l=[-90^\circ;-50^\circ]$. The boundaries in galactic latitude are not only driven by our choice of avoiding the Milky Way, but also by the limited sample of images available for WFC3 which had its first light in June 2009. The final calibrated, geometrically-corrected, dither-combined images with pixel scale 0.13\arcsec/pixel were inspected. They cover a total of 1.01\degr$^2$ of the sky.

In Fig.~\ref{fig:allsky} we show the distribution of the ACS and WFC3 images on the sky and  in Fig~\ref{fig:histogalaclat} we display the distribution of frames in galactic latitude. The combined ACS and WFC3 "surveys" totalize 
almost 7\degr$^2$ of imaging data all over the sky.

\begin{table}
   \caption{The WFC3 dataset. We give the number of frames inspected for each filter, as well as the number of lens candidates per filter. The last three columns give the minimum and maximum limiting magnitudes for the individual fields and the total area covered in each filter. The limiting magnitudes correspond to a 3-$\sigma$ detection in 1\arcsec$^2$.}
   \centering
   \begin{tabular}{c c c c c c c}
    \hline
    Filter& nb. frames & nb. lens & Min & Max & Total\\
     & & &mag.&mag.&area\\
    \hline
 G141&  38&0&--     &--& 0.05\\
F140W&  39&2&25.87&27.49&0.05\\
F105W&  61&1&24.76&27.00&0.08\\
F098M&  55&0&24.56&26.38&0.07\\
F110W&  80&1&25.08&26.85&0.10\\
F164N&  13&0&20.84&20.89&0.02\\
F139M&  74&0&24.06&24.68&0.10\\
F153M&   2&0&--    &--    &-- \\
F160W& 396&3&22.72&28.19&0.51\\
F125W& 112&2&25.29&27.59&0.14 \\
\hline
   \end{tabular}
   \label{tab:wfc3filter}
\end{table}


For each ACS image, we estimated the depth using the ACS Exposure Time Calculator for imaging (ETC\footnote{http://etc.stsci.edu/etc/input/acs/imaging/}), which is reliable for a very stable observatory such as the HST. This is also much less demanding in terms of computing time  than carrying out a full object detection on the actual data. 
 Our detection limits correspond to a total signal-to-noise ratio of 3 within an aperture of 1\arcsec$^2$ (3-$\sigma$ detection).

In Section~\ref{sec:compa} we compare the results  to that  of  COSMOS. This requires to measure the depth of our survey and that of COSMOS in the same way. Using the ACS ETC exactly in the same way for COSMOS and for our ACS F814W data we find a depth of $I=26.4$~mag for COSMOS, corresponding to an exposure time of 2028~s per pointing. Our estimate based on the ETC is well compatible with the limiting magnitudes quoted for COSMOS in other studies \citep[][]{Capak2007, Leauthaud2007}. 

\begin{figure}
\centering
\includegraphics[width= 0.5\textwidth]{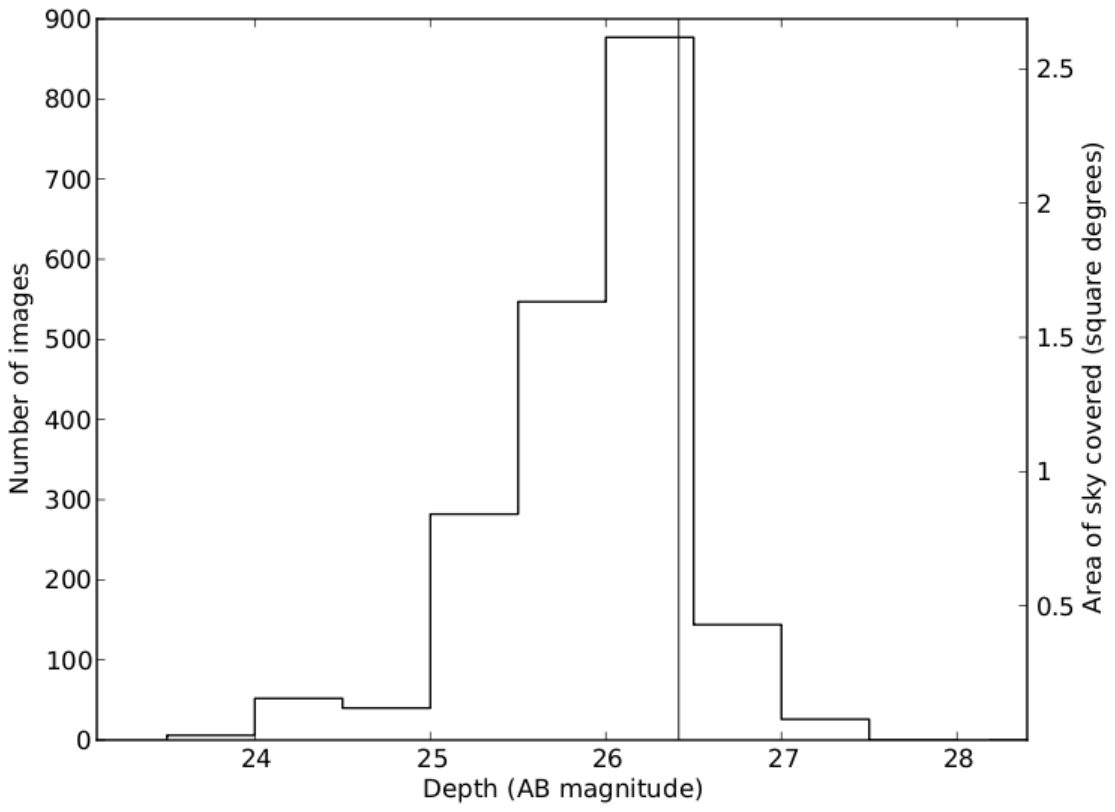}
\includegraphics[width= 0.5\textwidth]{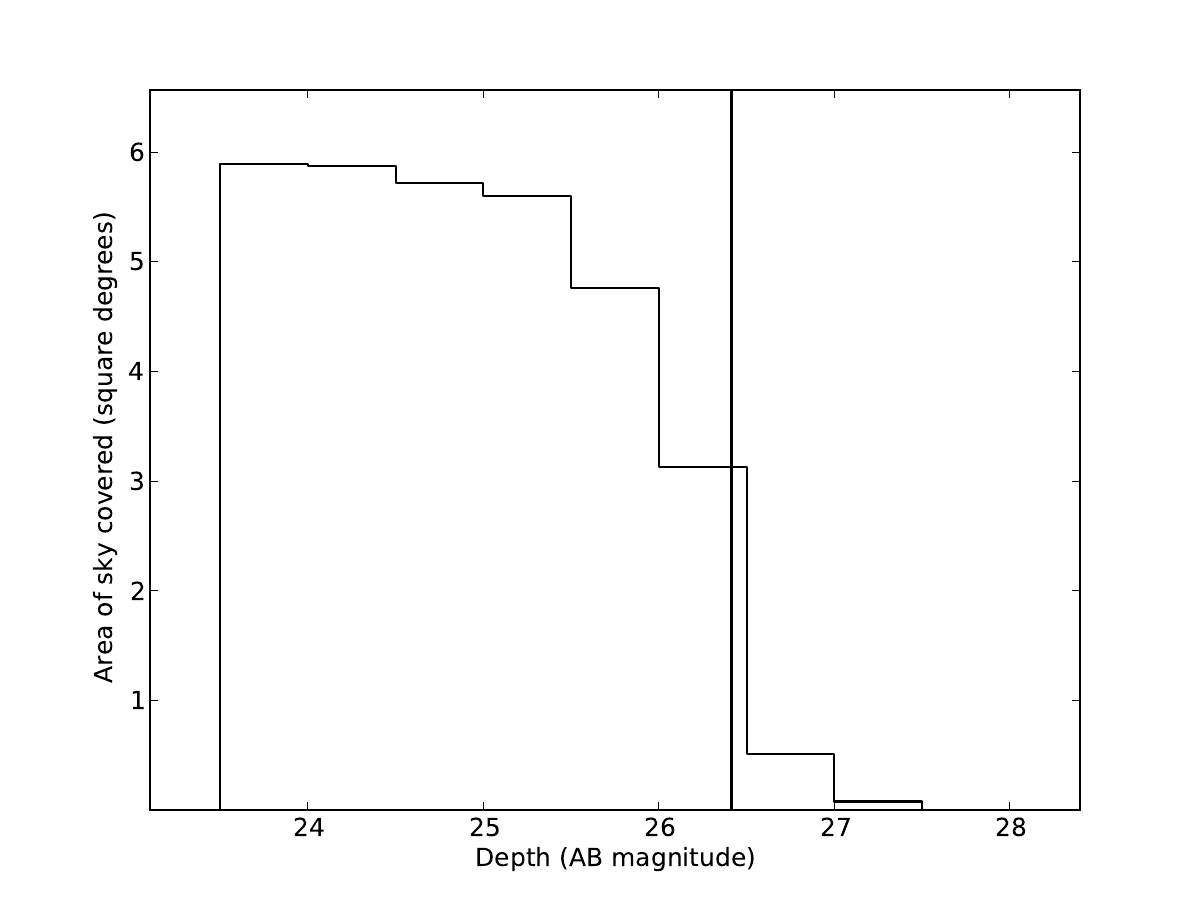}
\caption{{\it Top:}  The relative distribution of ACS image depths  observed in broadband I (F814W) filter. The corresponding area of sky covered by the images of a given depth is shown on the right axis.  The vertical line represents the depth of COSMOS survey. {\it Bottom:} The cumulative area of our survey as a function of depth.}
\label{fig:acsdepth}
\end{figure}

The ACS frames of our archival survey span a very broad range of exposure times, as shown in Fig.~\ref{fig:histoexptimeacs}. This translates into the distributions in depths displayed in Fig.~\ref{fig:acsdepth}, which peak in the bin [26.0-26.5]~mag, i.e. about the COSMOS depth (but the COSMOS data are not included in our search).  For comparison, in Fig.~\ref{fig:acsdepth}, the area covered by the  COSMOS field  would be 1.64\degr$^2$  for every depth up to $I=26.4$~mag.
 
\begin{table*}
\label{Table:1}
 
\caption{ List of lens candidates found from the HST/ACS data. Col.~1: Identifier.  Col.~2: Lens candidate name. Cols.~3 and 4: Coordinates. Col.~5: Radius of the arc in arc-second.  Col.~6: Total magnitude of the lensing galaxy in the F814W band, integrated in the 2-D light profile.
The typical errors are 1 to 2$\%$.
Col.~7: Effective radius in arcsec. The typical errors are 1 to 2$\%$.
Col.~8: Lensing galaxy axis ratio, q=($b/a$). The typical errors are 1 to 2$\%$.
 Col.~9: Lensing galaxy position angle, measured North to East. 
Col.~10: Comment, 3-digit code describing the lens candidate (see Section~\ref{class} for details). For candidate \#33 (ACS J153234.81+324115.3) no satisfactory fit was obtained. The unrealistic effective radius is not reported nor plotted in the Fig.~\ref{fig:radii}}
\begin{tabular}{llllllllll}
\hline
\vspace{0.5 mm}\\
ID&Name&RA&DEC&r$_{arc}$&I $($lens$)$&R$_{eff}$& q &PA&Comment\\
 & &h:m:s&$^{\circ}$:$\arcmin$:$\arcsec$ & $\arcsec$& F814W&$\arcsec$  &  &  &\\
\hline
\vspace{0.5 mm}\\
01&ACS J001423.02$-$302109.8&00:14:23.02&$-$30:21:09.8&1.52&18.86&0.25&0.70&$+$68.9&A1b\\  
02&ACS J001426.26$-$302255.9&00:14:26.26&$-$30:22:55.9&1.00&18.06&0.70&0.61&$+$60.0&A1a\\
03&ACS J005656.81$-$273841.0&00:56:56.81&$-$27:38:41.0&0.93&20.34&1.03&0.29&$-$102.9&A1b\\
04&ACS J010559.53$-$254438.0&01:05:59.53&$-$25:44:38.0&0.88&20.56&0.76&0.85&$+$34.9&A1b\\
05&ACS J011018.22+193819.5&01:10:18.22&+19:38:19.5 &3.10&17.03&2.47&0.90 &$+$33.1&A1a\\      
06&ACS J021634.62$-$051035.3&02:16:34.62&$-$05:10:35.3&1.69&18.28&0.57&0.44&$+$48.6&A1b\\
07&ACS J084710.65+344826.4&08:47:10.65&+34:48:26.4&1.98&20.25&0.66&0.88&$+$61.2&A1a\\
  &                       & 	      &           & &22.13&0.04&0.34&$+$158.6\\      
08&ACS J093323.88+441549.7&09:33:23.88&+44:15:49.7&1.32&19.71&0.26&0.93&$-$138.9&A1b\\
09&ACS J095139.44+684731.2&09:51:39.44&+68:47:31.2&0.77&19.78&0.50&0.71&$-$24.8&A1a\\
10&ACS J095718.69+685731.5&09:57:18.69&+68:57:31.5&0.82&18.62&1.29&0.91&$+$60.3&A1b\\       
11&ACS J103751.40$-$124327.5&10:37:51.40&$-$12:43:27.5&1.48&19.25&2.04&0.54&$+$106.6&A1b\\      
  &                         &           &          &  &19.99&0.68&0.54&$+$82.0\\      
12&ACS J104701.80+171739.7&10:47:01.80&+17:17:39.7& 0.46&19.34&0.86&0.72&$+$171.0& C2b\\      
13&ACS J111052.23+284233.7&11:10:52.23&+28:42:33.7&0.79&18.67&0.76&0.80&$-$56.2&A2c\\   
14&ACS J111127.06+285000.2&11:11:27.06&+28:50:00.2&0.39&21.02&1.50&0.69&$+$198.5&A1c\\      
15&ACS J111715.58+174500.0&11:17:15.58&+17:45:00.0&0.95&19.68&0.49&0.82&$+$152.4& C2b\\       
16&ACS J112310.02+141020.2&11:23:10.02&+14:10:20.2&1.01&20.50 &0.18&0.50&$+$102.2&A1c\\       
17&ACS J115135.05+004912.4&11:51:35.05&+00:49:12.4&1.87&19.10 &0.16&0.77&$-$60.8&A1c\\      
18&ACS J115138.59+004848.4&11:51:38.59&+00:48:48.4&1.96&19.12&1.24&0.41&$-$73.3&A1c\\      
19&ACS J121637.27$-$120042.2&12:16:37.27&$-$12:00:42.2&1.46&21.27&0.16&0.77&$+$51.0&A1b\\       
20&ACS J122332.64$-$123940.3&12:23:32.64&$-$12:39:40.3&0.36&19.68&0.19&0.54&$-$13.8&A1a\\      
21&ACS J130042.73+280523.3&13:00:42.73&+28:05:23.3&1.08&17.56&1.46&0.62&$-$112.5&A1b\\      
22&ACS J135401.12$-$123450.4&13:54:01.12&$-$12:34:50.4&2.95&19.32&1.35&0.83&$+$89.6&A1c\\       
23&ACS J140237.11+542716.4&14:02:37.11&+54:27:16.4&1.85&17.99&1.38&0.71&$-$114.4&A1a\\      
24&ACS J140339.94+541633.3&14:03:39.94&+54:16:33.3&0.90&18.66&3.44&0.34&$+$20.7&A1a\\      
25&ACS J141649.81+522951.7&14:16:49.81&+52:29:51.7&0.34&20.82&0.94&0.96&$-$42.9&A1b\\      
26&ACS J141657.72+433026.3&14:16:57.72&+43:30:26.3&0.78&20.82&0.16&0.79&$+$27.8&A1b\\      
27&ACS J141710.17+433133.2&14:17:10.17&+43:31:33.2&2.14&18.16&4.94&0.39&$+$123.0&A1a\\       
28&ACS J142028.48+525541.5&14:20:28.48&+52:55:41.5&0.61&19.67&0.56&0.91&$+$47.8&A1b\\       
29&ACS J142158.70+531045.9&14:21:58.70&+53:10:45.9&0.82&22.38&0.40&0.59&$-$164.9&A1c\\      
30&ACS J143542.20+342841.9&14:35:42.20&+34:28:41.9&0.96&19.96&1.77&0.43&$-$99.3&A1c\\      
31&ACS J144000.07+313123.5&14:40:00.07&+31:31:23.5&0.60&22.08&0.12&0.61&$+$142.8&A1b\\       
32&ACS J152702.08+355726.8&15:27:02.08&+35:57:26.8&1.73&19.59&1.16&0.68&$+$13.8&A1c\\       
33&ACS J153234.81+324115.3&15:32:34.81&+32:41:15.3&1.21&18.54&16.67&0.38&$+$123.4& C4b\\       
34&ACS J171750.68+593457.5&17:17:50.68&+59:34:57.5&1.79&17.92&0.68&0.77&$-$99.2&A1b\\      
35&ACS J171817.43+593146.4&17:18:17.43&+59:31:46.4&2.35&15.21&2.72&0.79&$-$49.7&A1a\\      
36&ACS J173923.28+690706.6&17:39:23.28&+69:07:06.6&1.23&20.94&0.14&0.38&$+$86.0&A1b\\      
37&ACS J174035.58+690328.1&17:40:35.58&+69:03:28.1&1.12&19.08&0.63&0.31&$+$86.7&A1a\\      
38&ACS J221501.12$-$135822.9&22:15:01.12&$-$13:58:22.9&0.82&19.10 &0.78&0.22&$-$168.9&A1a\\      
39&ACS J230325.68+085212.6&23:03:25.68&+08:52:12.6&0.72&17.62&1.08&0.35&$-$108.9&A1a\\      
40&ACS J235130.60$-$261459.7&23:51:30.60&$-$26:14:59.7&3.32&16.26&2.49&0.74&$-$125.2&A1a\\      
\hline
\end{tabular}
\label{tab:acscan}
\end{table*}


\begin{table*}
\label{Table:2}
 \caption[]{List of lens candidates found from our HST/WFC3 data. The description of the columns is the same as for Table~\ref{tab:acscan} except that the magnitudes are for the filter given in the "Filter" column. 
\label{tab:WFC3can}
}
\begin{tabular}{lllllllllll}
\hline
\vspace{0.5 mm}\\
ID&Name&RA&DEC&r$_{arc}$&R$_{eff}$& q &PA&Mag&Filter&Comment\\
 & &h:m:s&$^{\circ}$:$\arcmin$:$\arcsec$ &$\arcsec$ &$\arcsec$&& &\\
\hline
\vspace{0.5 mm}\\
1&WFC3 J023924.56$-$013600.82&02:39:24.56&$-$01:36:00.82&0.91&0.65&0.60&$+$111.9&20.50&F140W&R1b\\
2&WFC3 J033227.44$-$275520.42&03:32:27.44&$-$27:55:20.42&1.83&1.85&0.57&$+$145.5&18.35&F105W&A1a\\
3&WFC3 J033245.17$-$274940.46&03:32:45.17&$-$27:49:40.46&1.71&0.25&0.11&$+$47.5&20.48&F125W&A1a\\
4&WFC3 J100023.53+021653.34&10:00:23.53&+02:16:53.34&2.08&1.29&0.50&$-$73.2&19.33&F140W&A1b\\
5&WFC3 J140221.89+094520.16&14:02:21.89&+09:45:20.16&1.04&3.67&0.78&$-$99.4&20.13&F110W&A2c\\
6&WFC3 J142030.50+530249.23&14:20:30.50&+53:02:49.23&0.73&0.35&0.57&$-$101.3&22.84&F125W&A1c\\
7&WFC3 J143703.21+350153.55&14:37:03.21&+35:01:53.55&0.69&0.55&0.64&$+$61.6&19.87&F160W&R1a\\
8&WFC3 J171336.64+585640.73&17:13:36.64&+58:56:40.73&1.43&5.98&0.87&$-$14.4&19.07&F160W&A1c\\
9&WFC3 J171736.34+601437.60&17:17:36.34&+60:14:37.60&2.23&0.23&0.69&$-$92.2&20.69&F160W&A1b\\
\hline
\end{tabular}
\end{table*}

\begin{table}
\label{Table:3}
\caption{Filters used to create the colour images in Fig.~\ref{fig:ACScolor}.   When only two filters are available, the mean of the two images was taken as the green channel.}
\label{tab:colorfilters}
\begin{tabular}{llll}
\hline
Name& Red & Green & Blue\\
\hline
ACS J001423.02$-$302109.8&F814W & F606W & F435W\\  
ACS J001426.26$-$302255.9&F814W & F606W& F435W\\
ACS J021634.62$-$051035.3&F814W&\_&F606W\\
ACS J095139.44+684731.2&  F814W&\_& F606W\\
ACS J104701.80+171739.7&  F814W&\_& F435W\\      
ACS J111052.23+284233.7&  F814W&\_&F555W\\     
ACS J115135.05+004912.4&  F814W&\_&F435W\\      
ACS J130042.73+280523.3&  F814W&\_&F475W\\      
ACS J140237.11+542716.4&  F814W& F555W& F435W\\      
ACS J140339.94+541633.3&  F814W& F555W& F435W\\      
ACS J141649.81+522951.7&  F814W&\_&F606W\\      
ACS J141657.72+433026.3&  F814W&\_&F606W\\      
ACS J141710.17+433133.2&  F814W&\_& F606W\\       
ACS J142028.48+525541.5&  F814W&\_&F606W\\       
ACS J142158.70+531045.9&  F814W&\_& F606W\\      
ACS J144000.07+313123.5&  F814W&\_& F606W\\       
ACS J230325.68+085212.6&  F814W&\_&F435W\\      

\hline
\end{tabular}
\end{table}

\section{The lens catalogue}

\subsection{The search by visual inspection}\label{subsec:inspection}

The HST images were inspected  by two of the authors: RSP for the ACS images and RK for the WFC3 images. We use for this purpose a simple {\tt Python} 
interface designed to ease the image display with custom intensity levels and scales. The purpose of the interface is to allow us explore quickly the full dynamic range of the images. This is particularly useful to detect rings and arcs in the inner parts of galaxies and groups of galaxies, even without subtraction of the foreground light.
This tool divides the images into rectangular grid of regions of 500 $\times$ 500 pixels which are inspected one at a time with 100\% zoom and
generates {\tt jpeg and fits} cut-outs  images of the candidates. The coordinates of any candidate are obtained by interfacing with the Aladin Sky Atlas\footnote{http://aladin.u-strasbg.fr/}.
This simple tool allows us to inspect seven to ten  HST frames per hour.

Some of the fields we use are pointed observations of known gravitational lenses. We identify those as being parts of previous lens catalogues including BELLS \citep{2012brownstein}, CASTLE\footnote{http://www.cfa.harvard.edu/castles/}, SL2S \citep{Cabanac2007}, SLACS \citep{2006ApJ...638..703B, 2008ApJ...682..964B} and SWELLS \citep{2011MNRAS.417.1601T}. We do not include these lenses in our catalogue, as they are not the result of a blind search, but we use the corresponding images for our blind search for new lenses. 
Note that CASTLE and SL2S contain very few ACS/F814W images and that the combined SLACS, SWELLS and BELLS samples cover only 0.3\degr$^2$ of the sky, i.e. 4\% of our surveyed area. Clustering of lenses in the field of already known strong lenses \citep[e.g.][]{Fassnacht2006} is therefore unlikely to bias significantly our sample towards artificially large numbers of lenses.

\begin{figure*}F
\fbox{
\includegraphics[width= 0.48\textwidth]{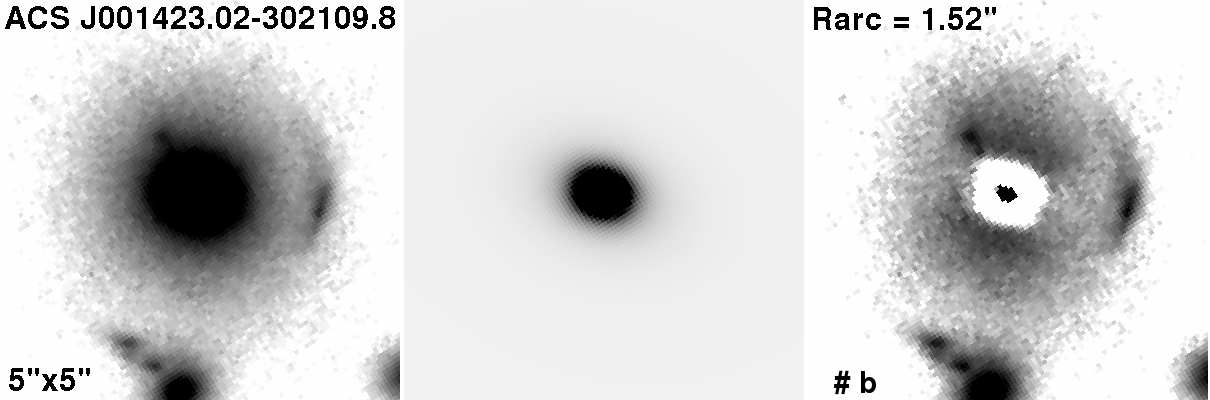}}
\fbox{
\includegraphics[width= 0.48\textwidth]{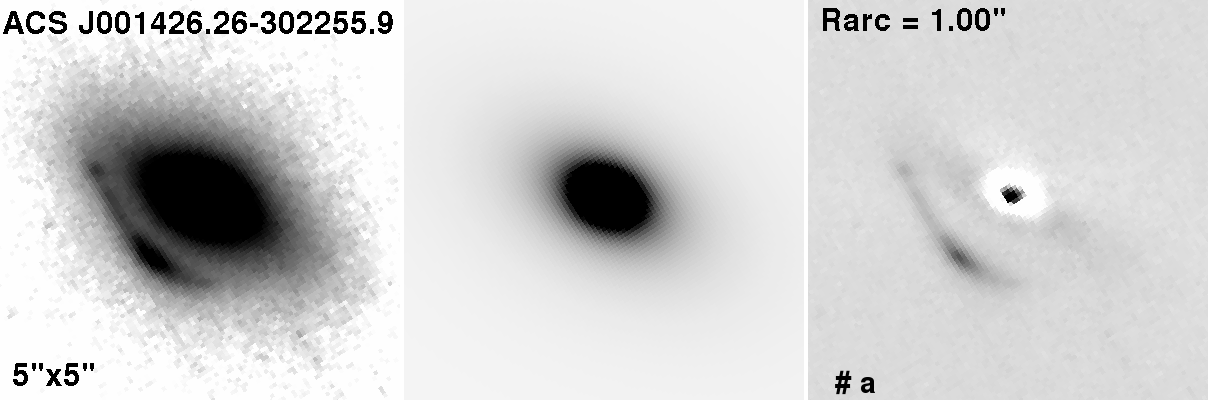}}
\fbox{
\includegraphics[width= 0.48\textwidth]{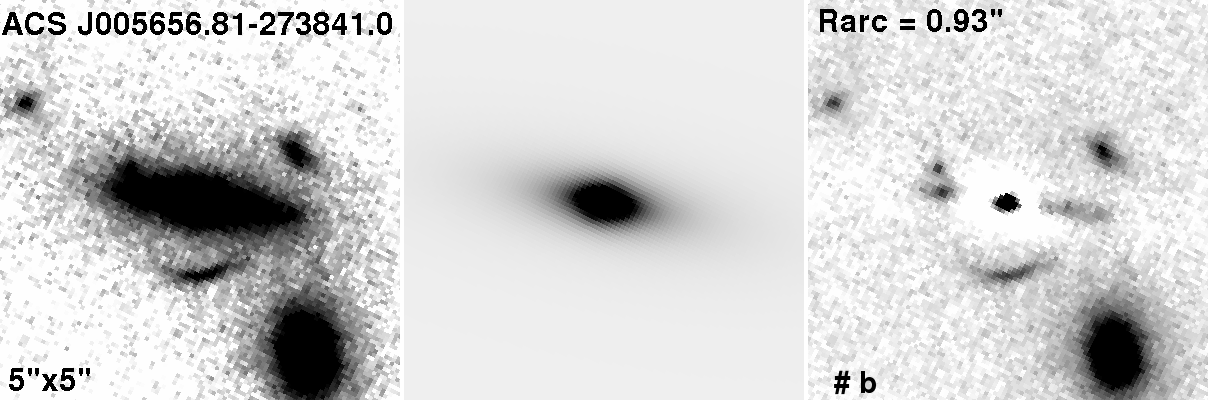}}
\fbox{
\includegraphics[width= 0.48\textwidth]{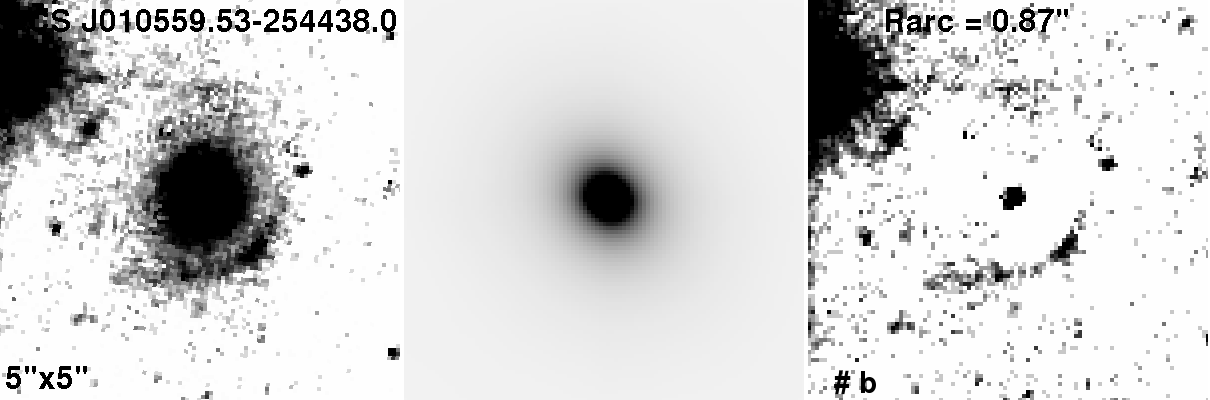}}
\fbox{
\includegraphics[width= 0.48\textwidth]{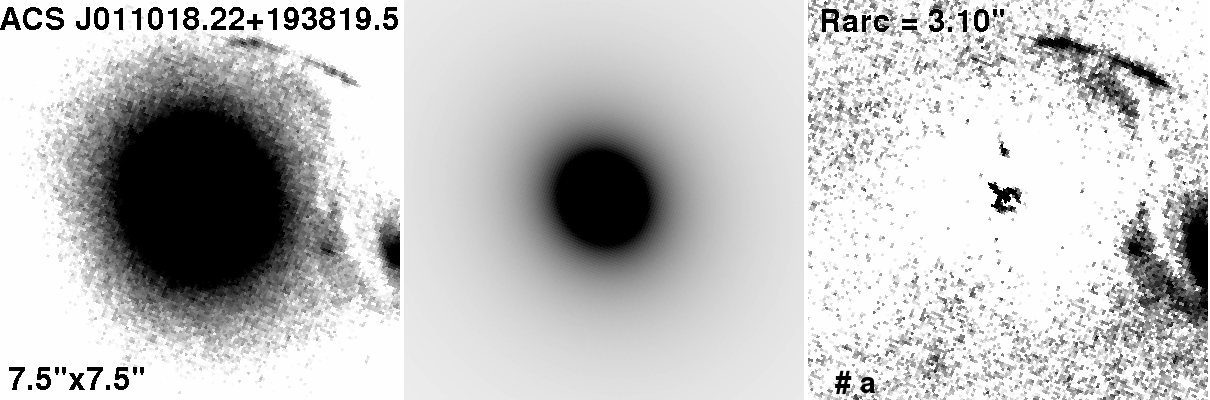}}
\fbox{
\includegraphics[width= 0.48\textwidth]{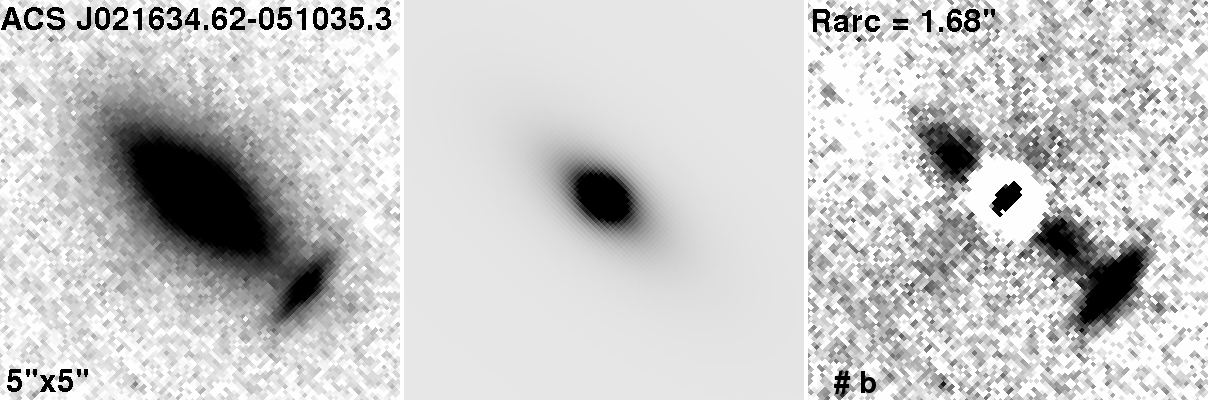}}
\fbox{
\includegraphics[width= 0.48\textwidth]{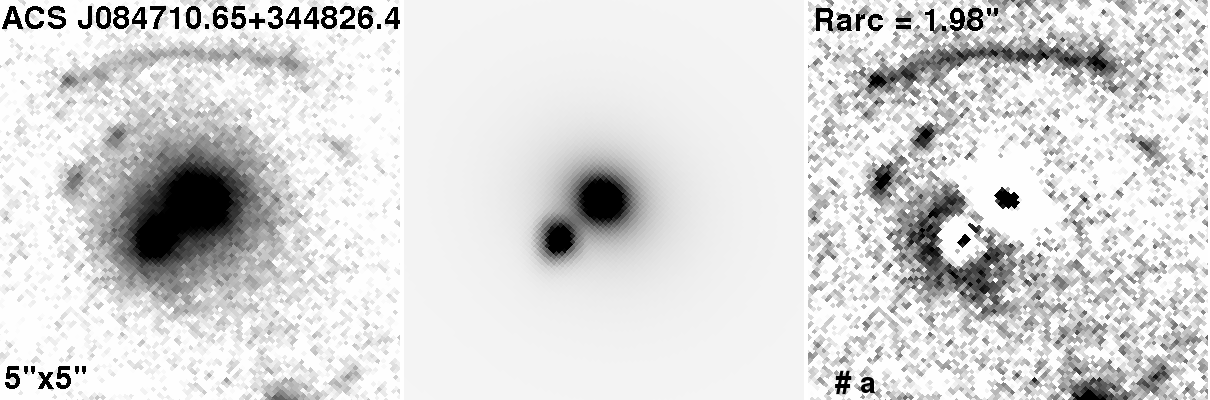}}
\fbox{
\includegraphics[width= 0.48\textwidth]{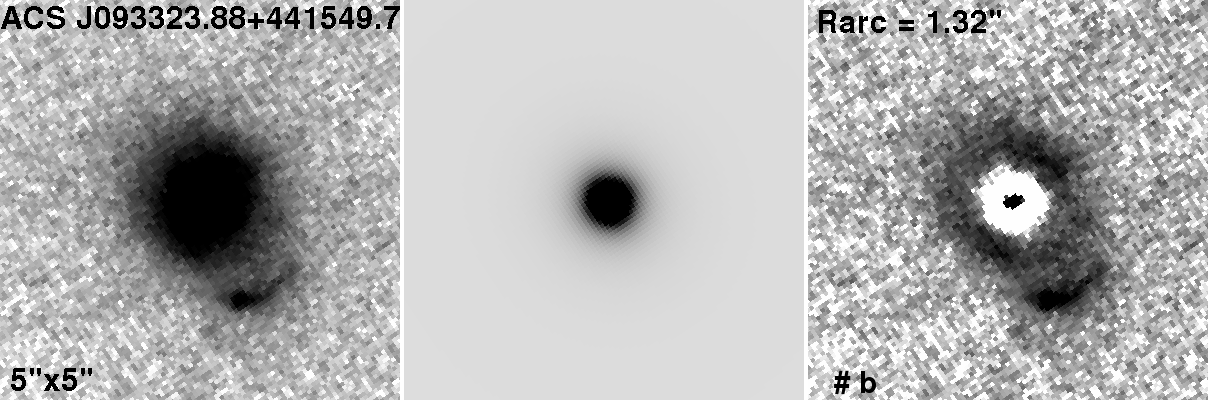}}
\fbox{
\includegraphics[width= 0.48\textwidth]{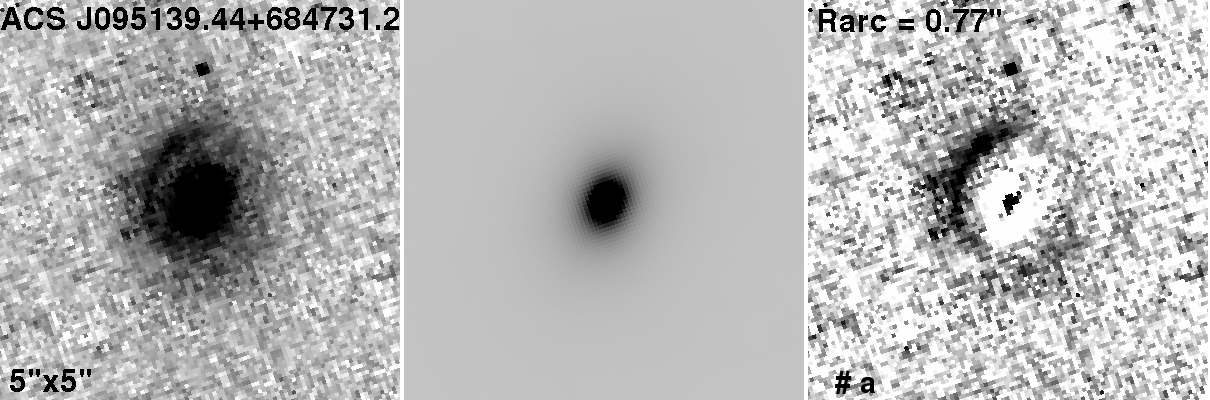}}
\fbox{
\includegraphics[width= 0.48\textwidth]{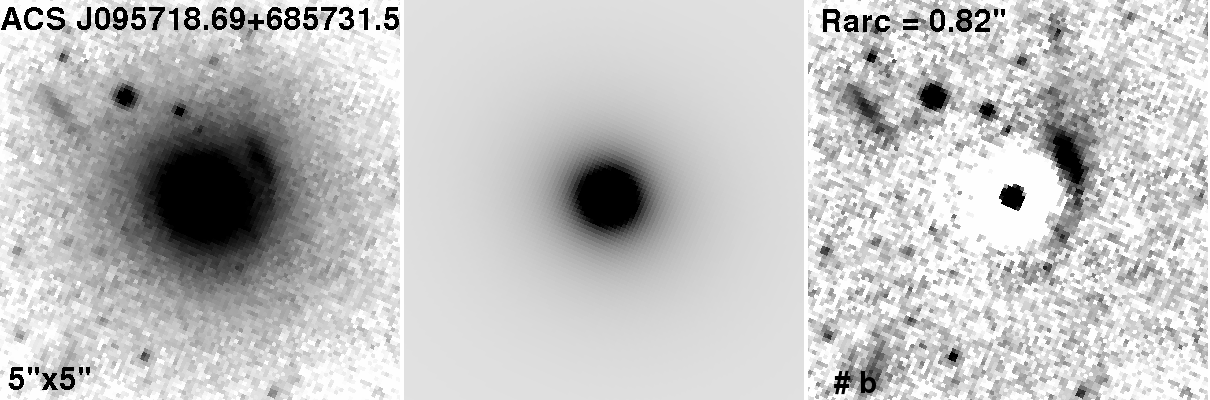}}
\fbox{
\includegraphics[width= 0.48\textwidth]{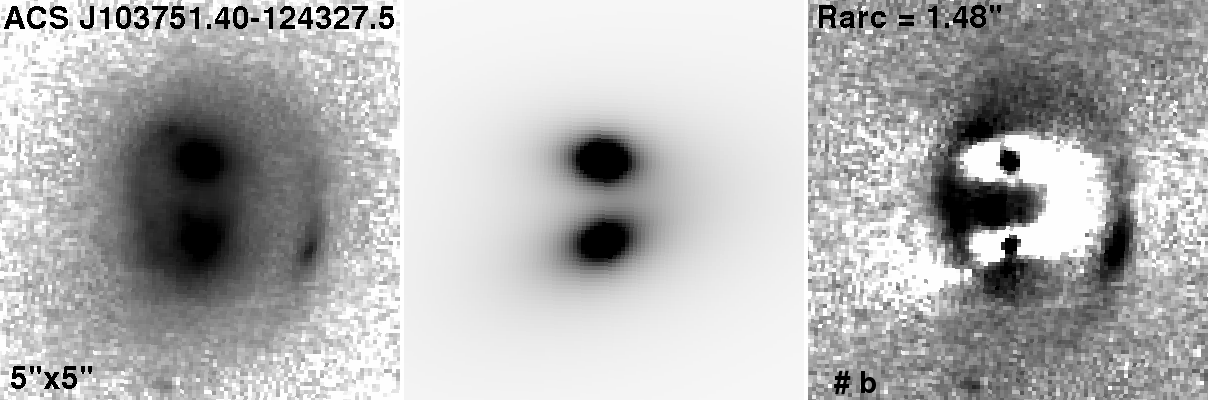}}
\fbox{
\includegraphics[width= 0.48\textwidth]{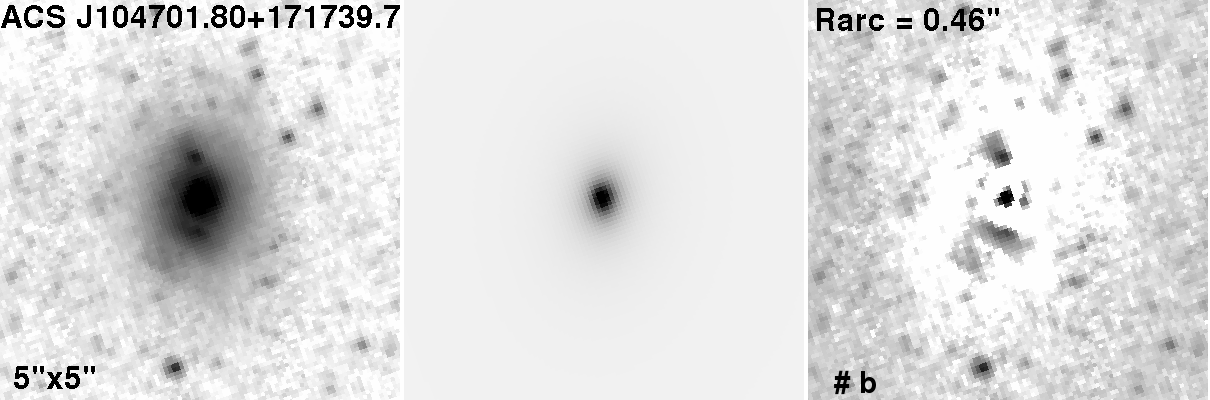}}
\fbox{
\includegraphics[width= 0.48\textwidth]{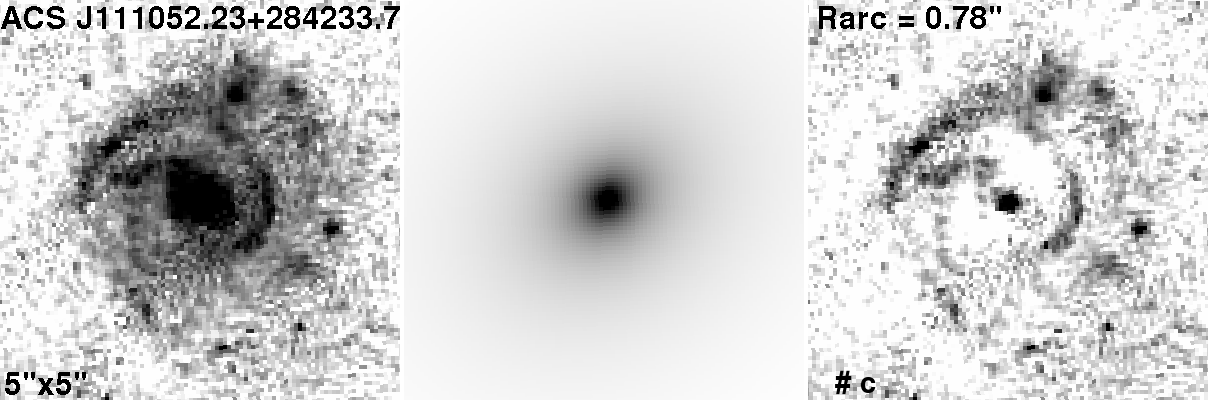}}
\fbox{
\includegraphics[width= 0.48\textwidth]{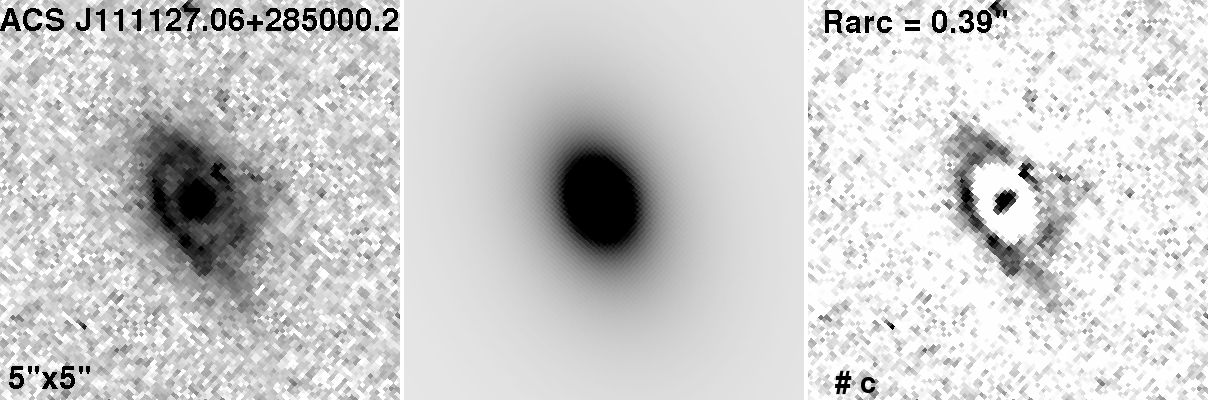}}
\caption{Lens candidates found by visual inspection of the ACS images. There are two lens candidates per line. For both are displayed  in three panels : the ACS image, the 2-D light profile for the lens galaxy using GALFIT, and the image where the model is subtracted from the original data. Orientation of the images is north is to the top and East is to the left. The letter in the lower-left corner of the residual image is the confidence level given to each system to be a lens.} 
\label{fig:acscompo1}
\end{figure*}
\begin{figure*}
\fbox{
\includegraphics[width= 0.48\textwidth]{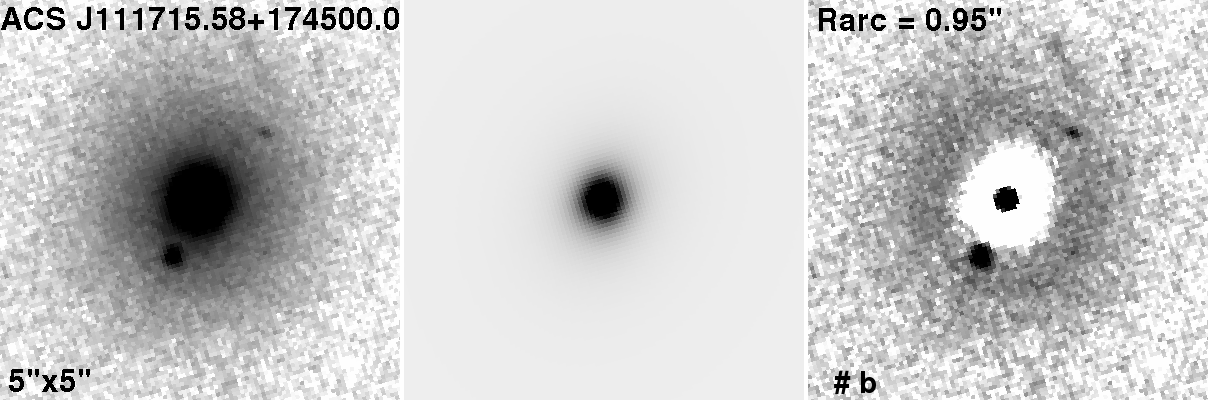}}
\fbox{
\includegraphics[width= 0.48\textwidth]{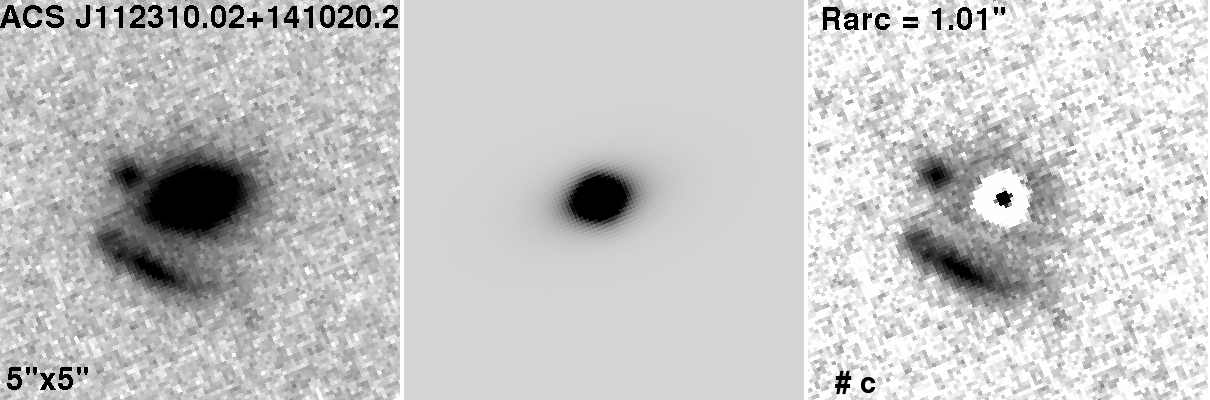}}
\fbox{
\includegraphics[width= 0.48\textwidth]{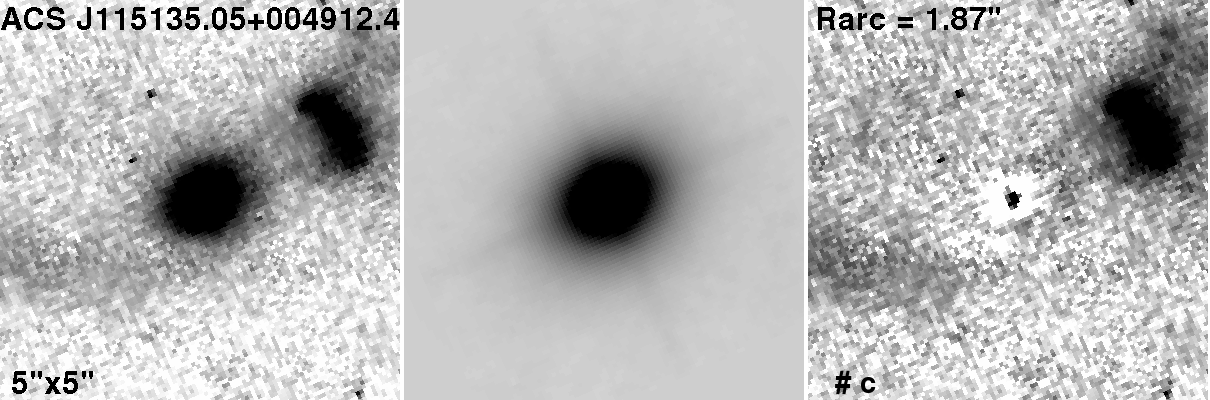}}
\fbox{
\includegraphics[width= 0.48\textwidth]{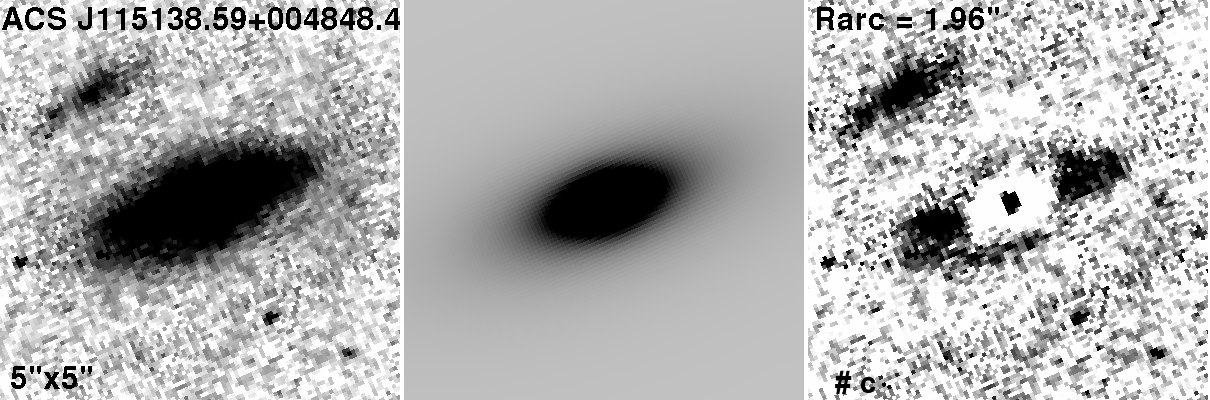}}
\fbox{
\includegraphics[width= 0.48\textwidth]{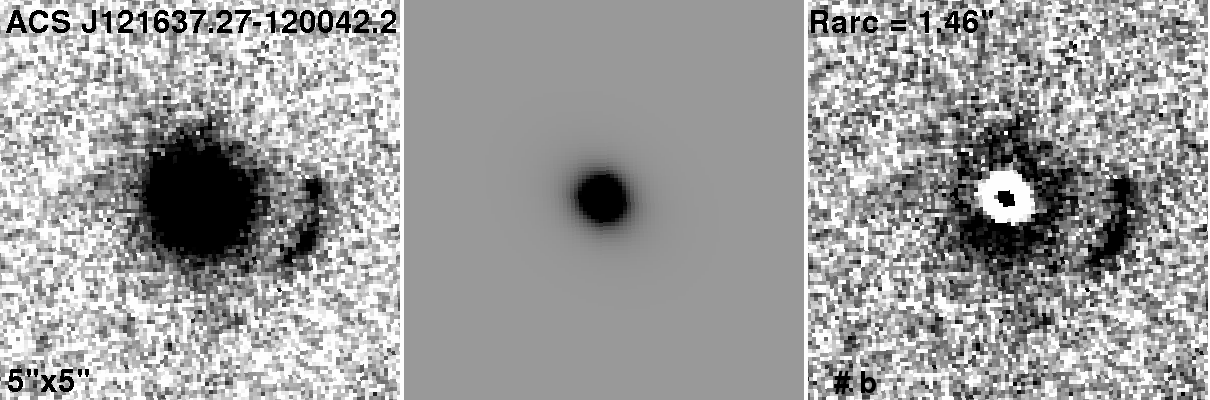}}
\fbox{
\includegraphics[width= 0.48\textwidth]{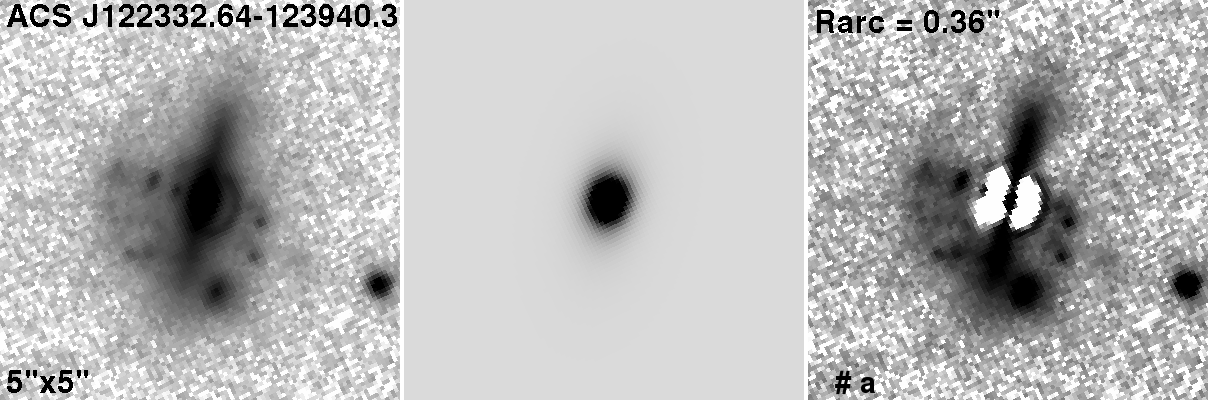}}
\fbox{
\includegraphics[width= 0.48\textwidth]{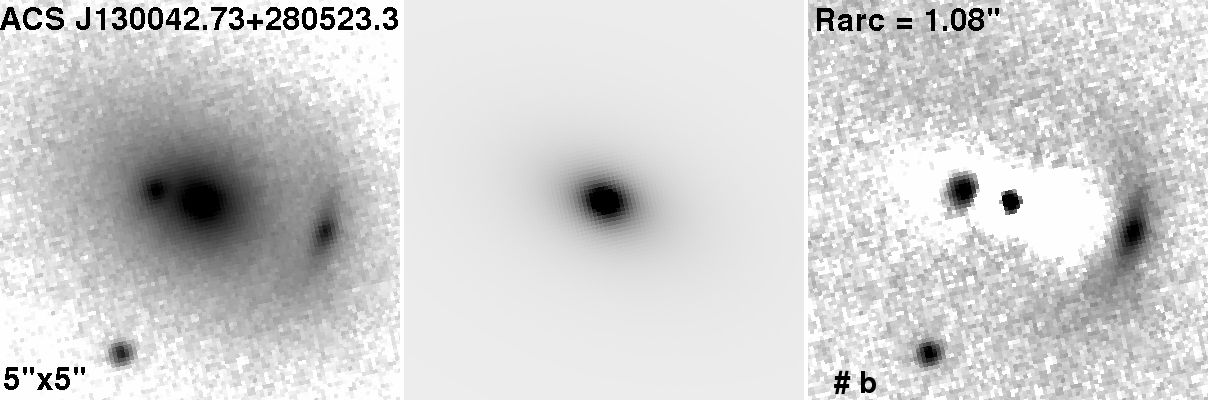}}
\fbox{
\includegraphics[width= 0.48\textwidth]{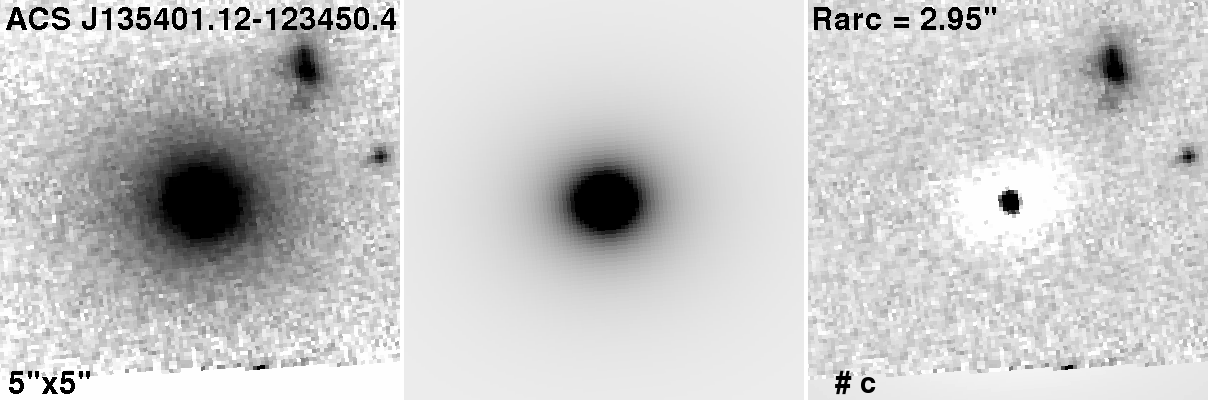}}
\fbox{
\includegraphics[width= 0.48\textwidth]{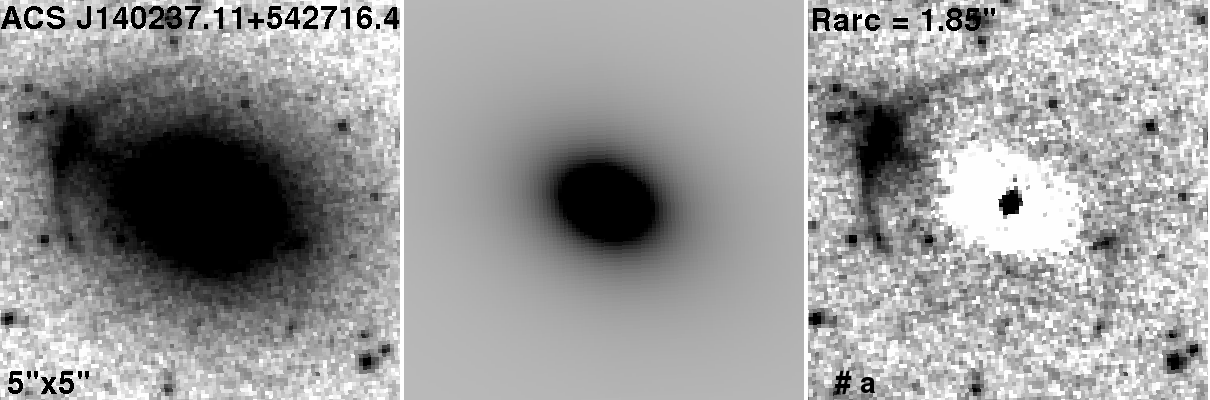}}
\fbox{
\includegraphics[width= 0.48\textwidth]{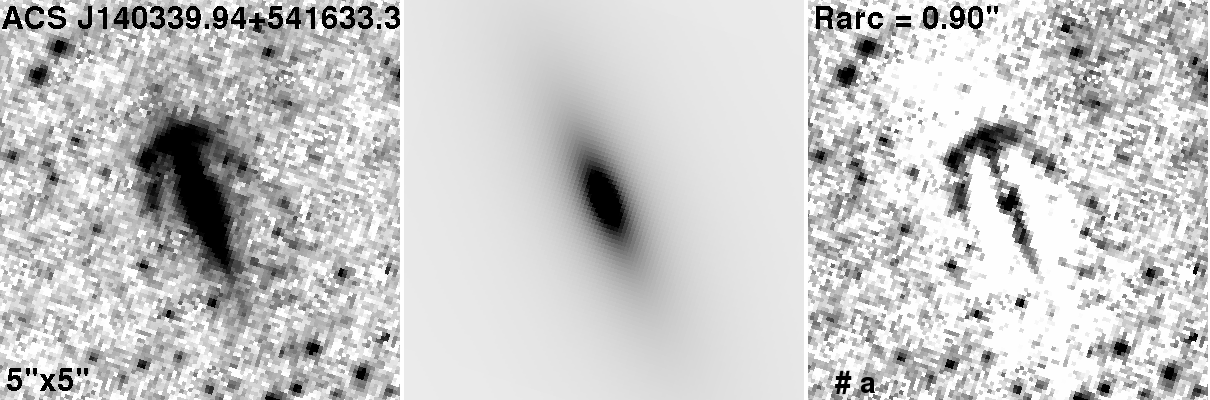}}
\fbox{
\includegraphics[width= 0.48\textwidth]{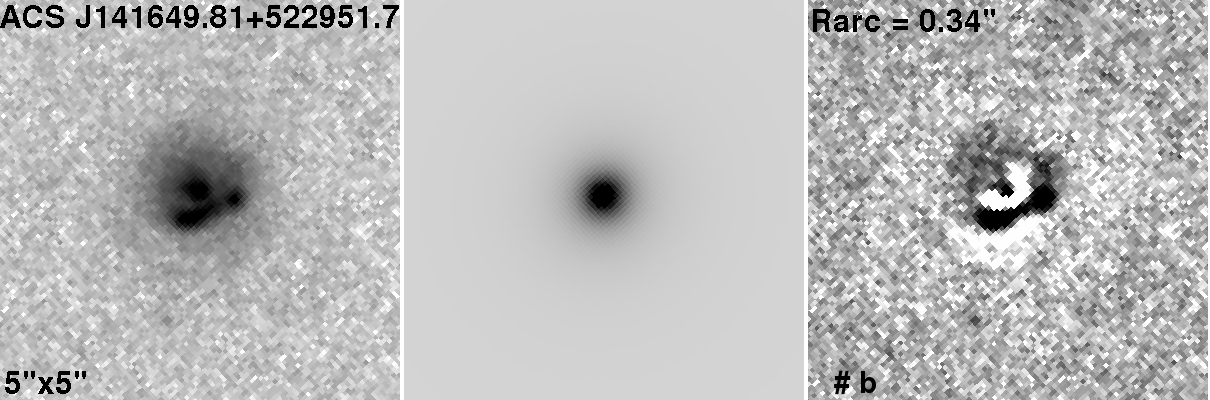}}
\fbox{
\includegraphics[width= 0.48\textwidth]{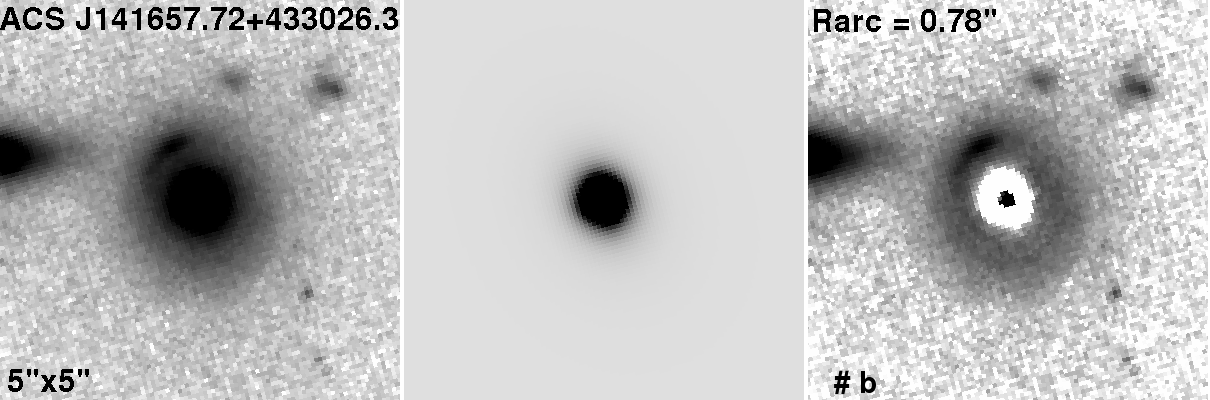}}
\fbox{
\includegraphics[width= 0.48\textwidth]{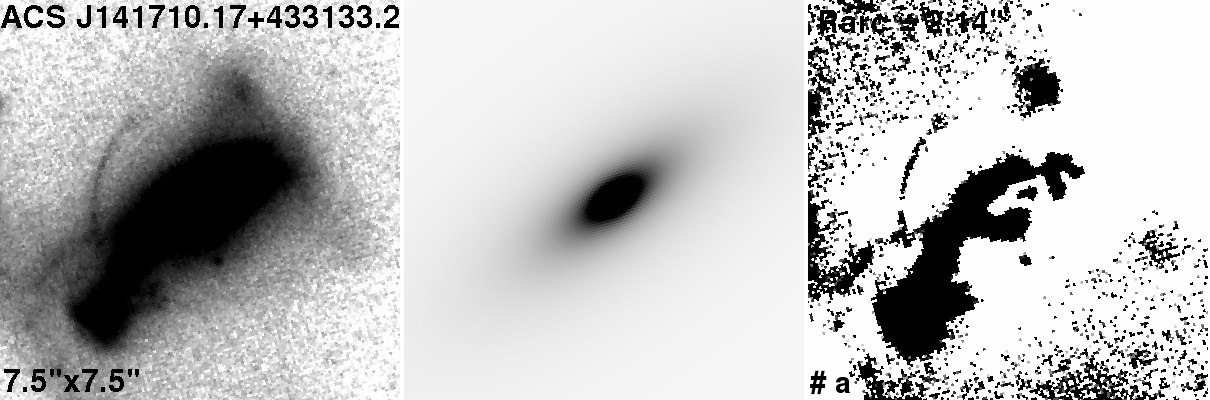}}
\fbox{
\includegraphics[width= 0.48\textwidth]{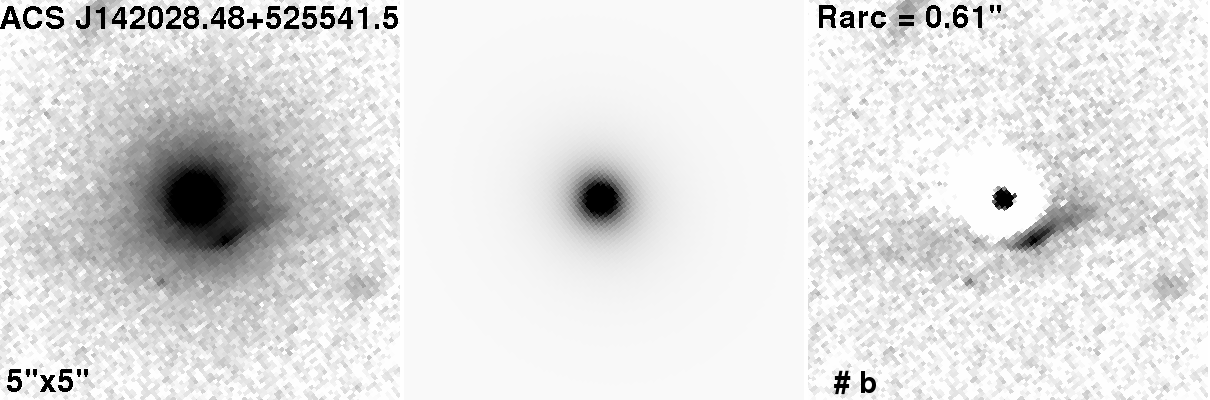}}
\caption{Continue Fig.~\ref{fig:acscompo1}.}
\label{fig:acscompo2}
\end{figure*}
\begin{figure*}
\fbox{
\includegraphics[width= 0.48\textwidth]{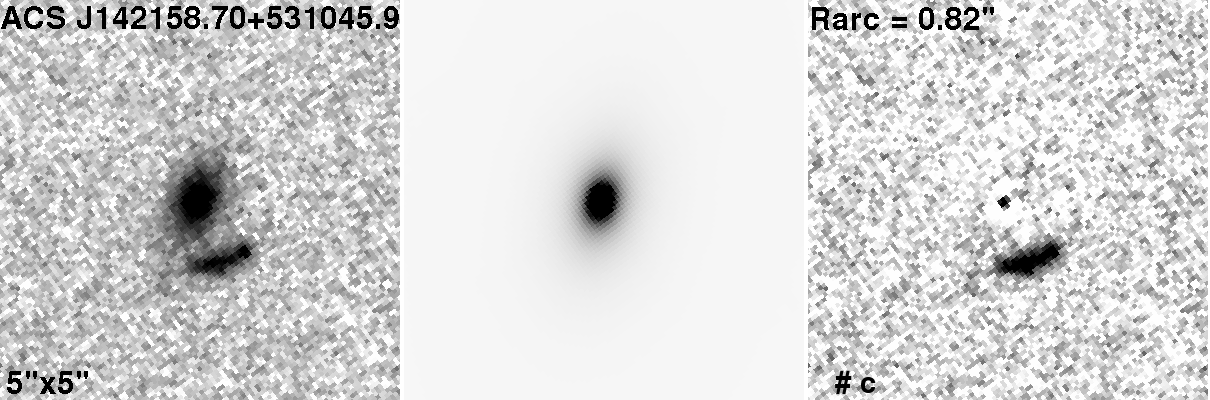}}
\fbox{
\includegraphics[width= 0.48\textwidth]{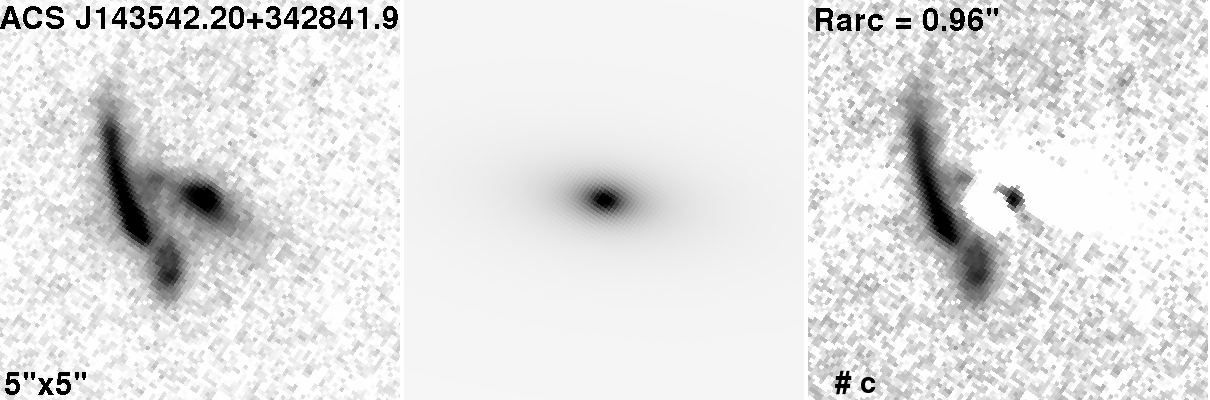}}
\fbox{
\includegraphics[width= 0.48\textwidth]{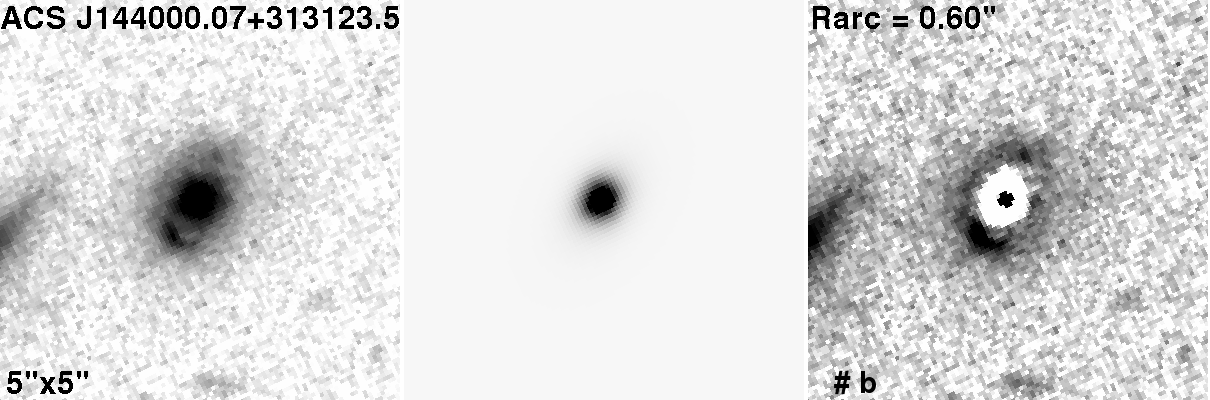}}
\fbox{
\includegraphics[width= 0.48\textwidth]{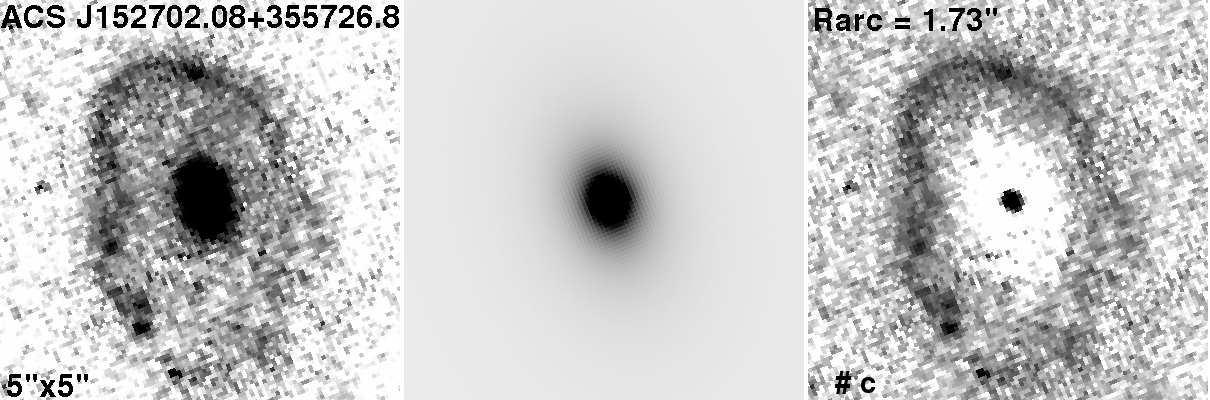}}
\fbox{
\includegraphics[width= 0.48\textwidth]{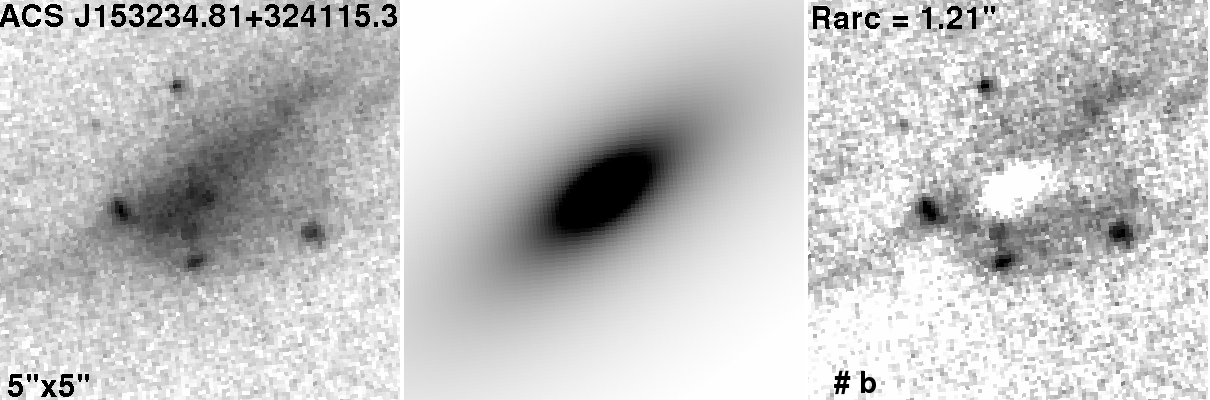}}
\fbox{
\includegraphics[width= 0.48\textwidth]{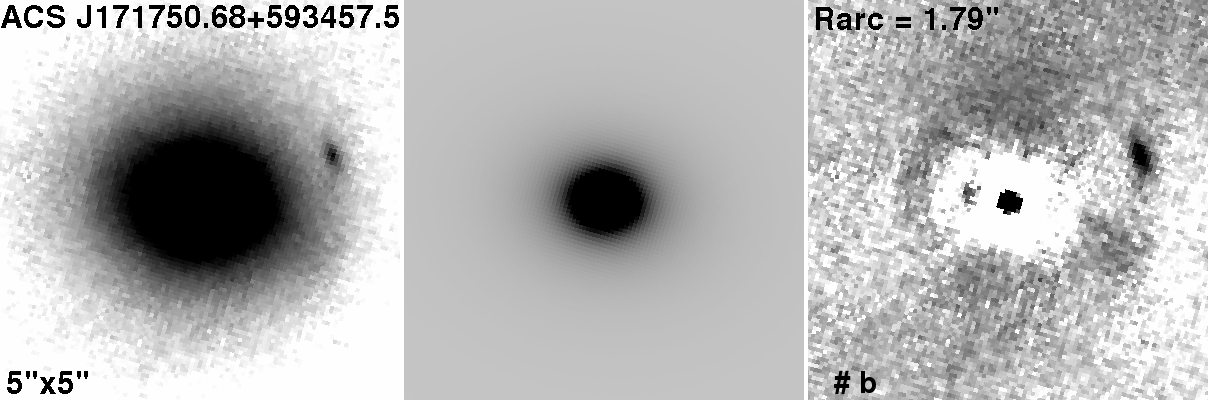}}
\fbox{
\includegraphics[width= 0.48\textwidth]{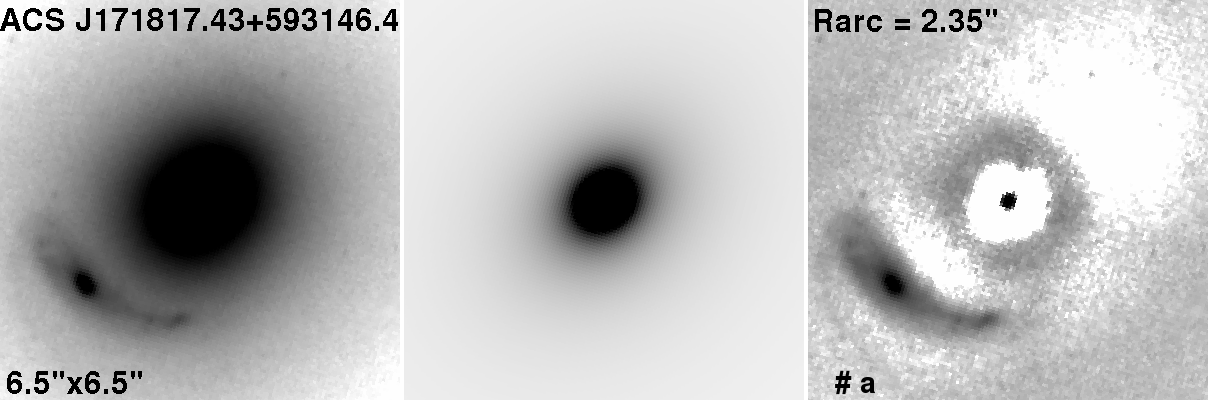}}
\fbox{
\includegraphics[width= 0.48\textwidth]{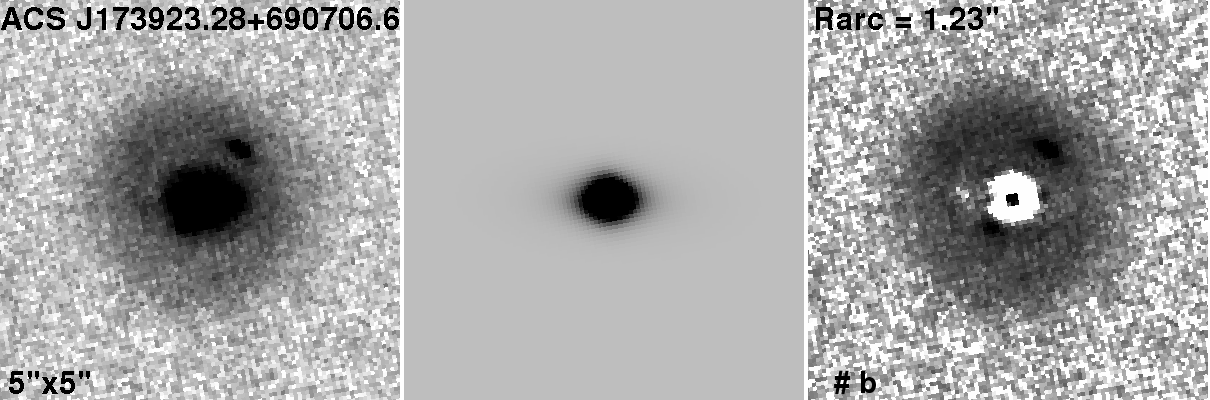}}
\fbox{
\includegraphics[width= 0.48\textwidth]{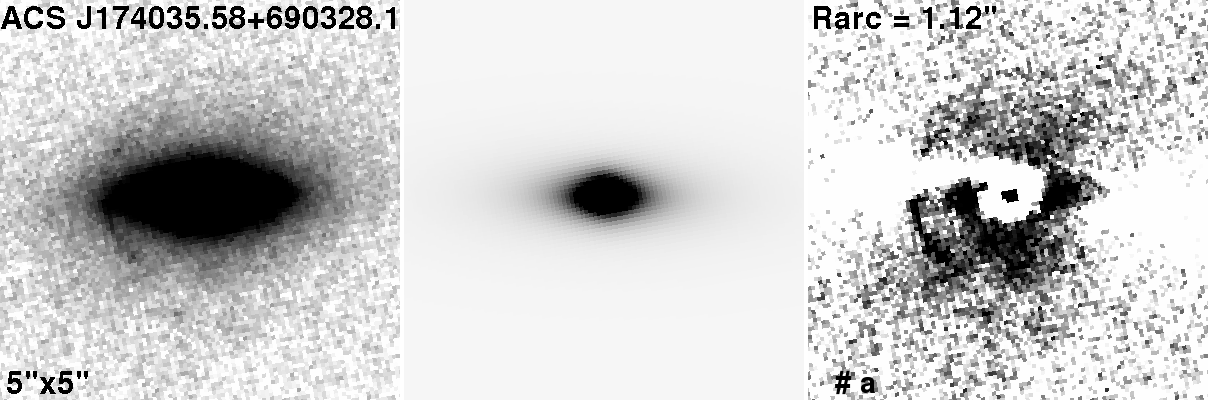}}
\fbox{
\includegraphics[width= 0.48\textwidth]{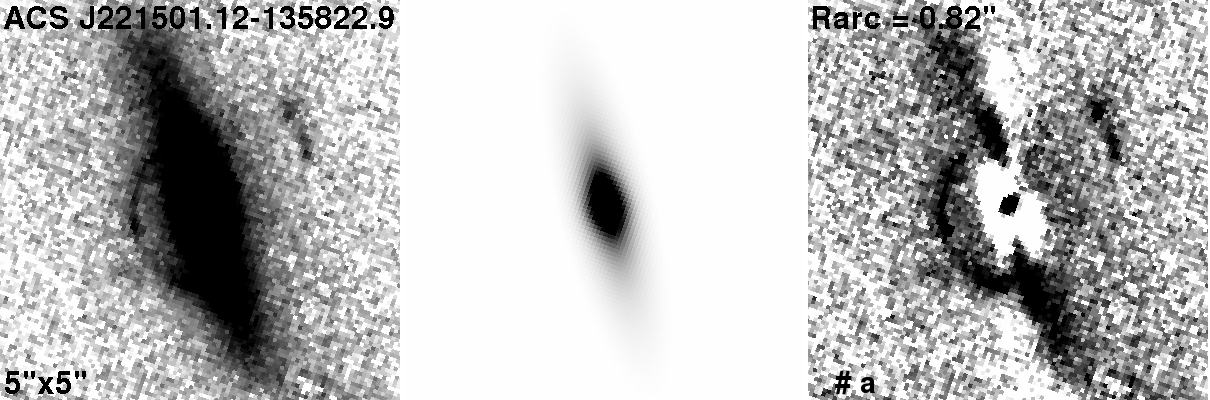}}
\fbox{
\includegraphics[width= 0.48\textwidth]{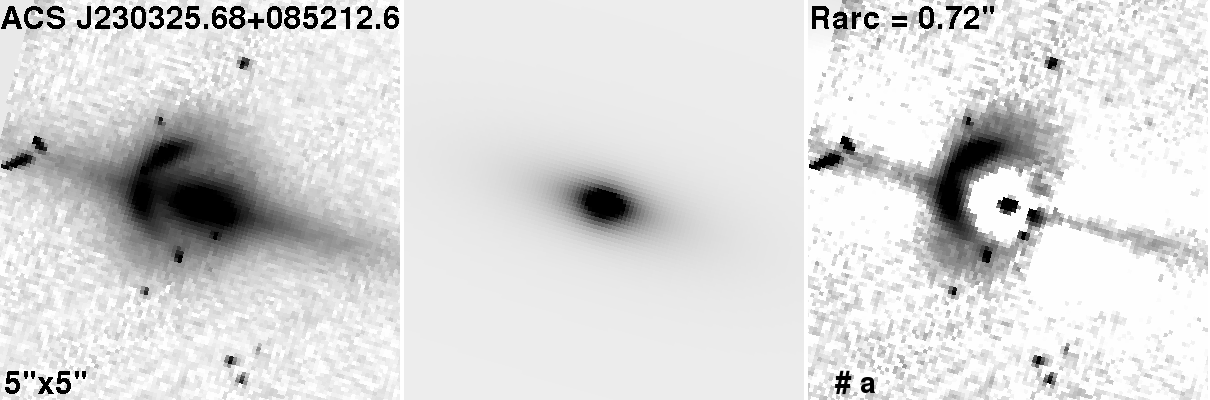}}
\fbox{
\includegraphics[width= 0.48\textwidth]{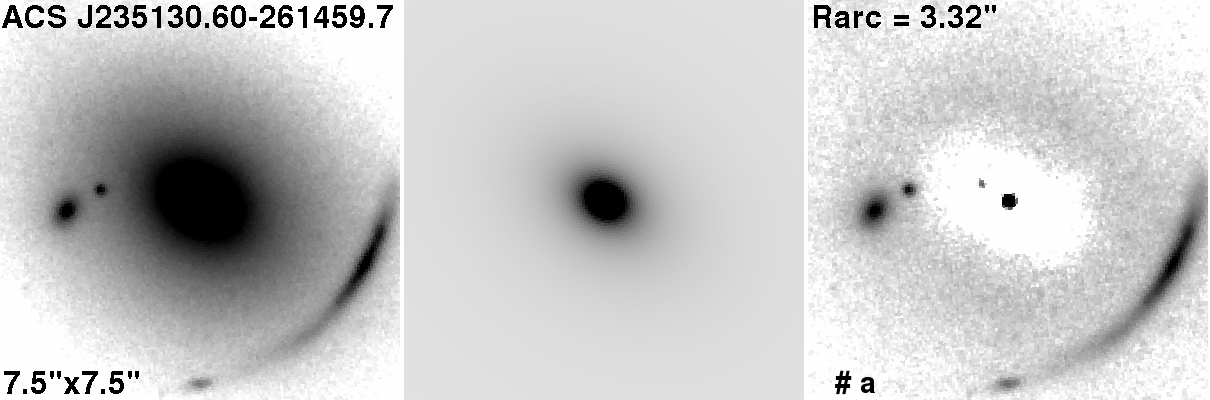}}
\caption{Continue Fig.~\ref{fig:acscompo1}.}             
\label{fig:acscompo3}
\end{figure*}
\begin{figure*}
\fbox{
\includegraphics[width= 0.48\textwidth]{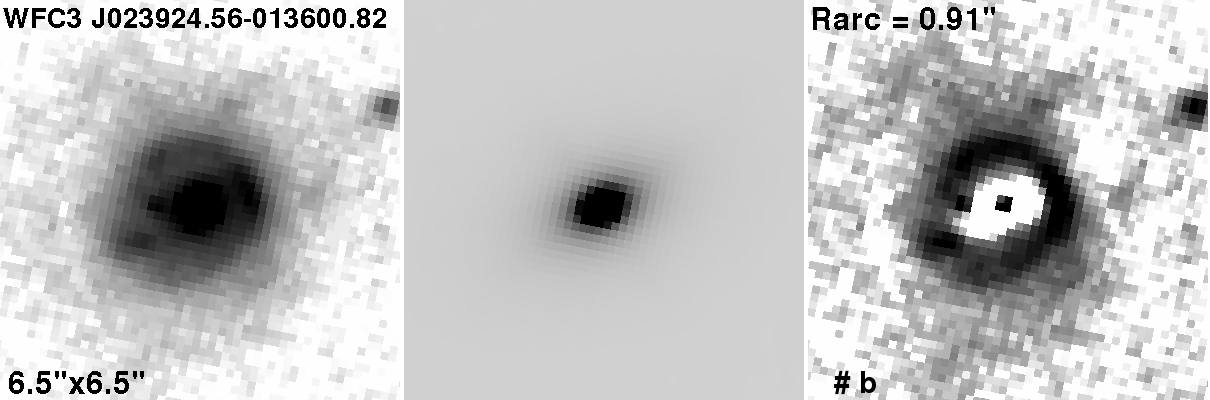}}
\fbox{
\includegraphics[width= 0.48\textwidth]{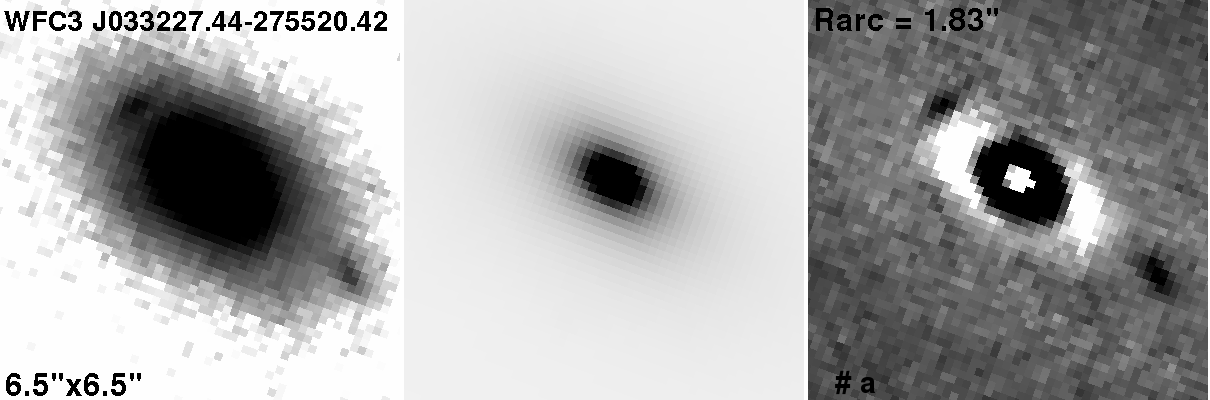}}
\fbox{
\includegraphics[width= 0.48\textwidth]{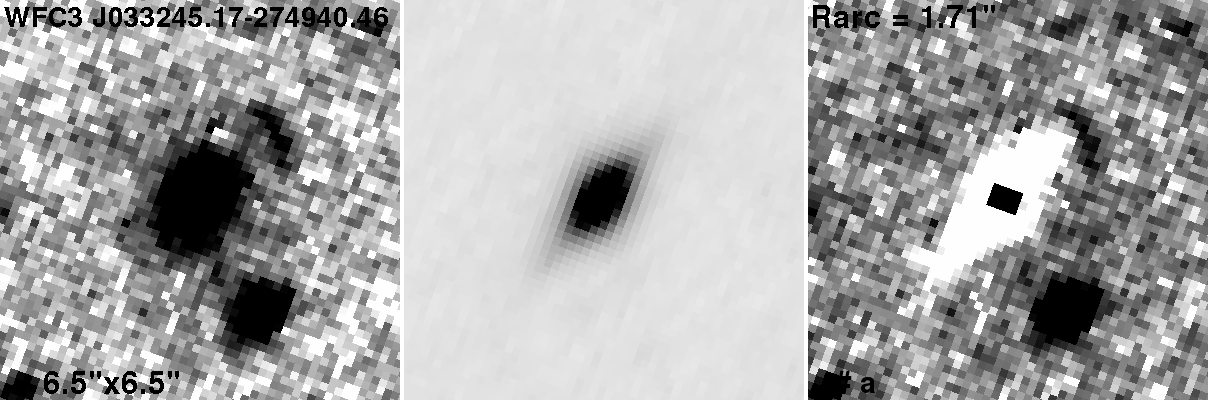}}
\fbox{
\includegraphics[width= 0.48\textwidth]{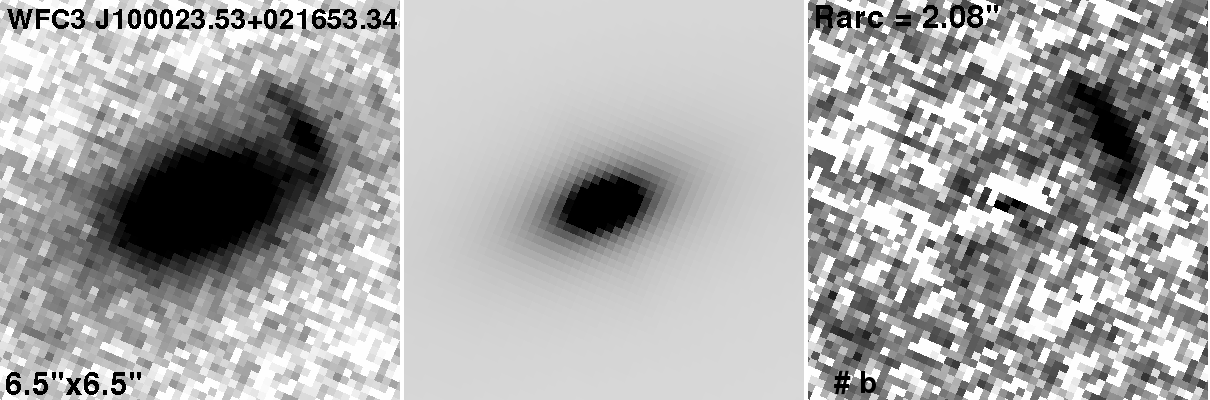}}
\fbox{
\includegraphics[width= 0.48\textwidth]{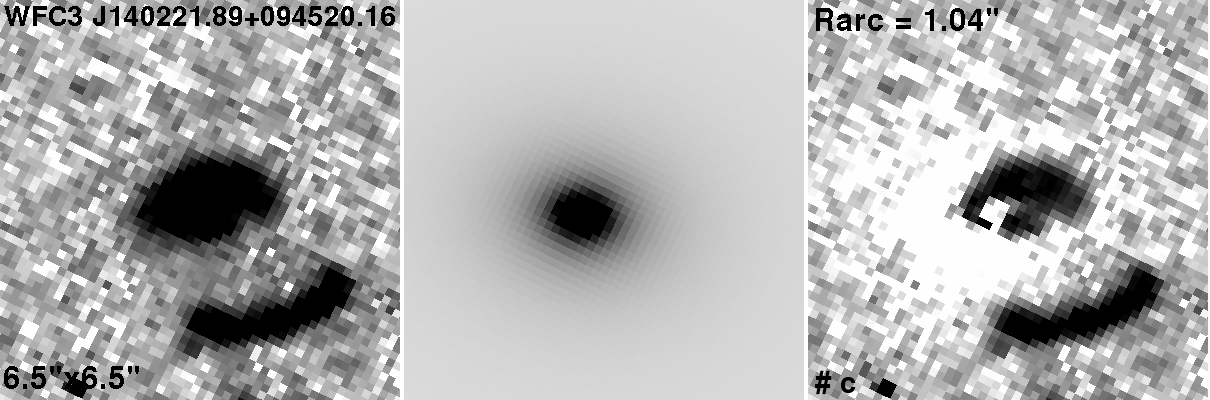}}
\fbox{
\includegraphics[width= 0.48\textwidth]{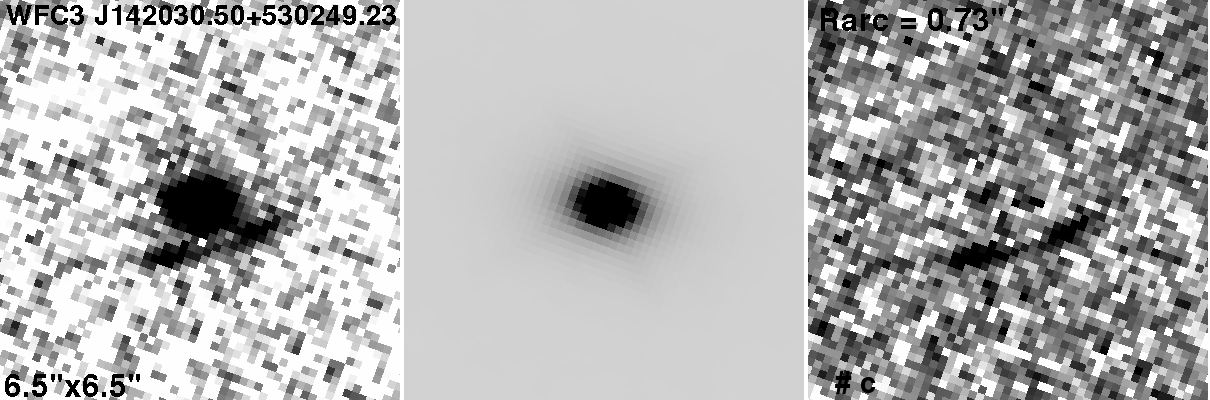}}
\fbox{
\includegraphics[width= 0.48\textwidth]{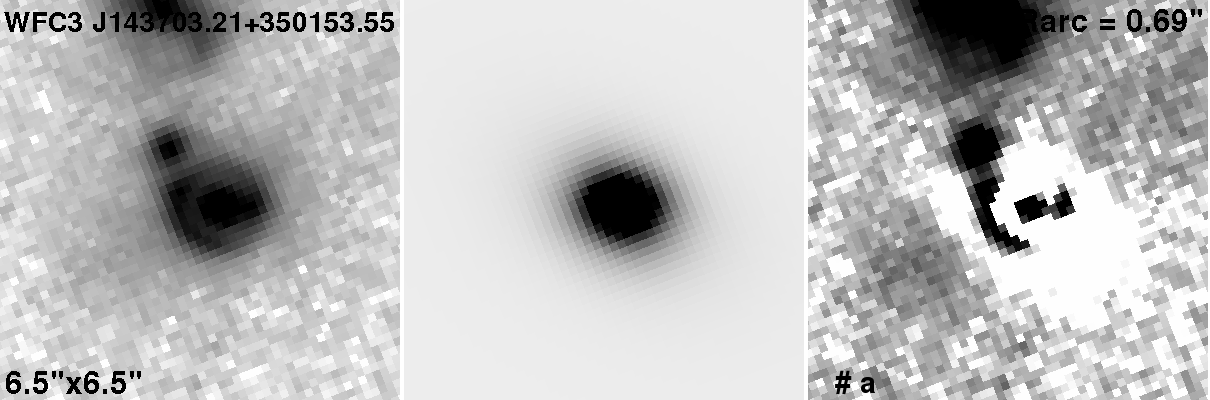}}
\fbox{
\includegraphics[width= 0.48\textwidth]{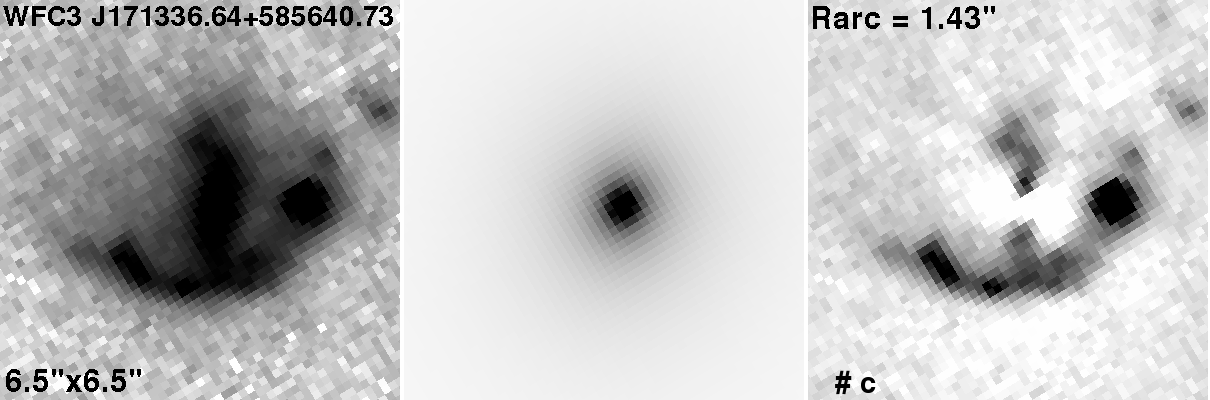}}
\fbox{
\includegraphics[width= 0.48\textwidth]{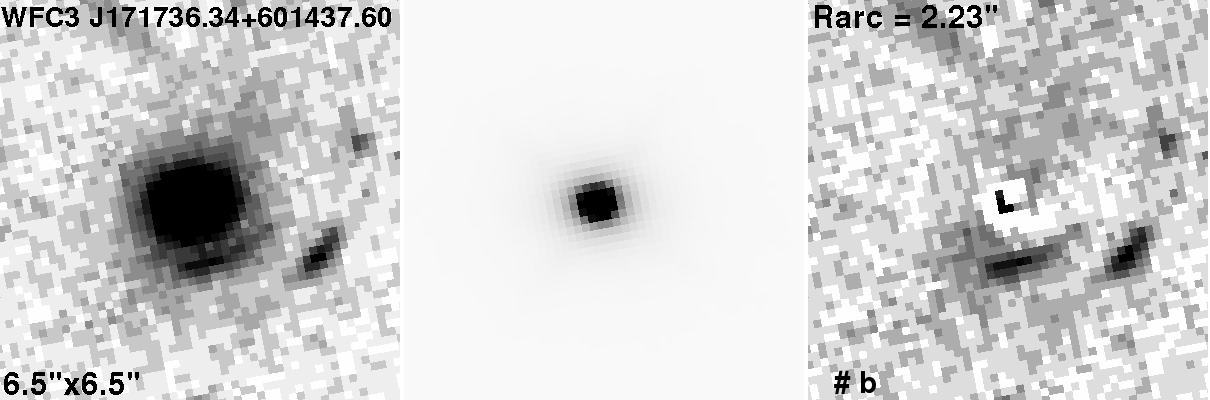}}
\caption{Lens candidates found by visual inspection of the WFC3 near-IR images.  There are two lens candidates per line. For both are displayed  in three panels : the WFC3 image, the 2-D light profile for the lens galaxy using GALFIT, and the image where the model is subtracted from the original data. Orientation of the images is north is to the top and East is to the left.}
\label{fig:WFC3compo1}
\end{figure*}
\begin{figure*}
\includegraphics[width= 4.2cm]{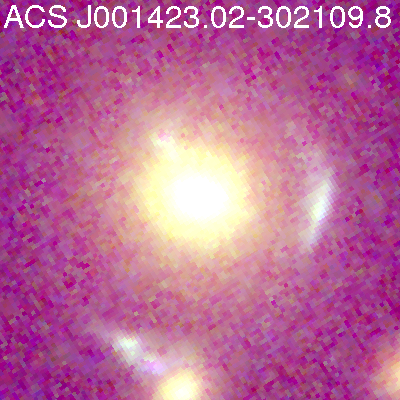}\includegraphics[width= 4.2cm]{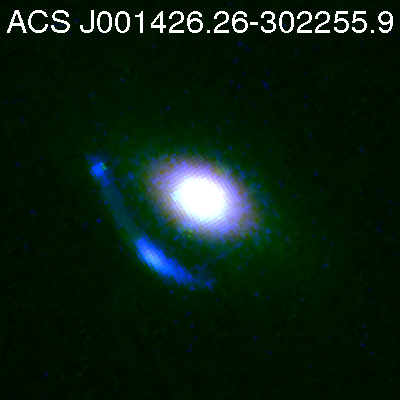}\includegraphics[width= 4.2cm]{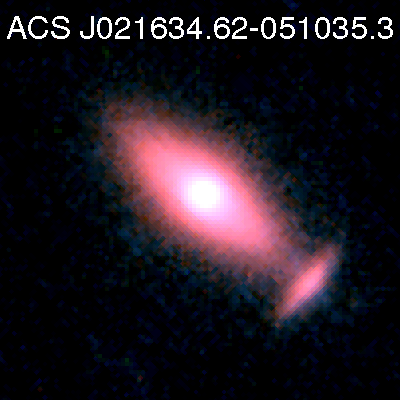}\includegraphics[width= 4.2cm]{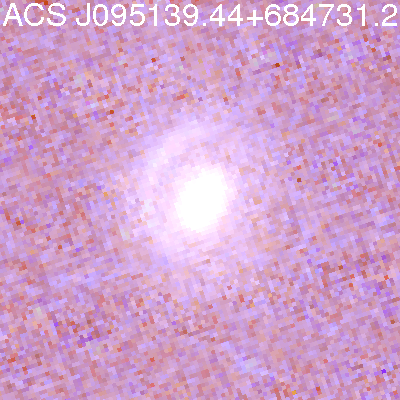}
\includegraphics[width= 4.2cm]{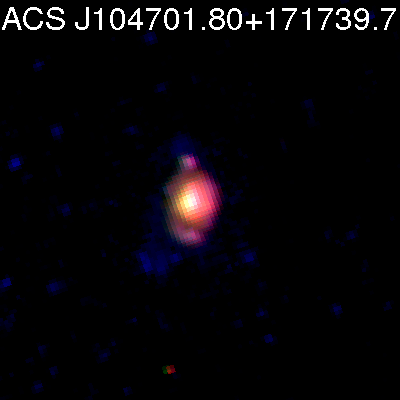}\includegraphics[width= 4.2cm]{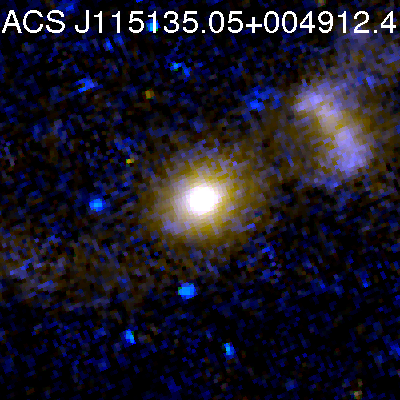}\includegraphics[width= 4.2cm]{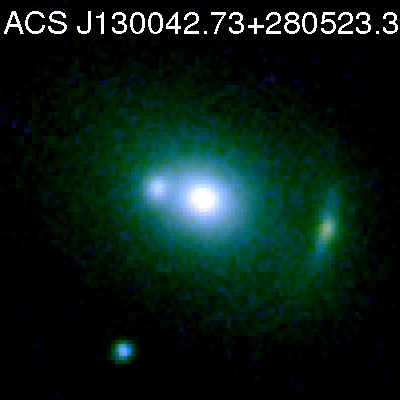}\includegraphics[width= 4.2cm]{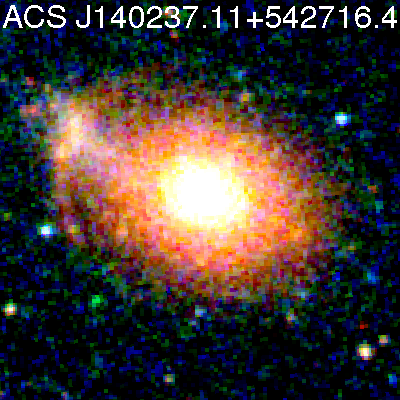}
\includegraphics[width= 4.2cm]{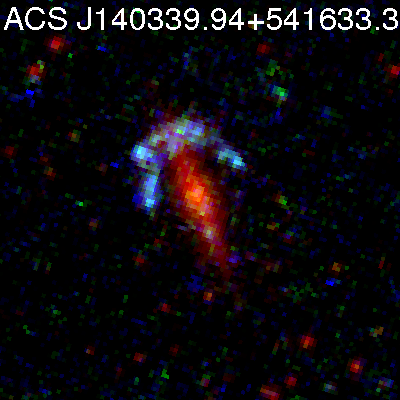}\includegraphics[width= 4.2cm]{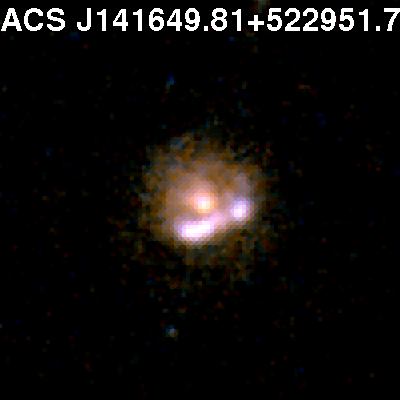}\includegraphics[width= 4.2cm]{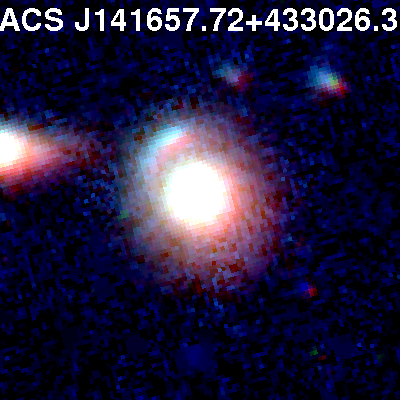}\includegraphics[width= 4.2cm]{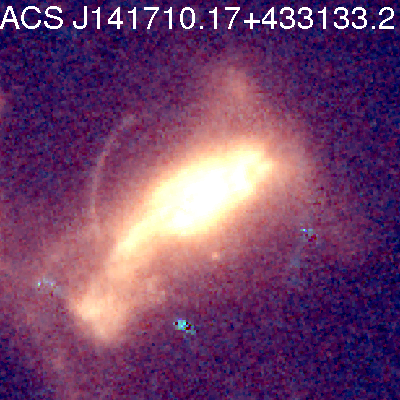}
\includegraphics[width= 4.2cm]{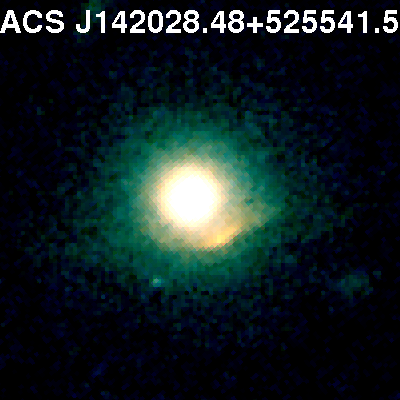}\includegraphics[width= 4.2cm]{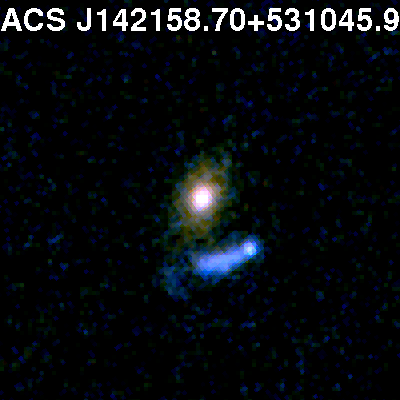}\includegraphics[width= 4.2cm]{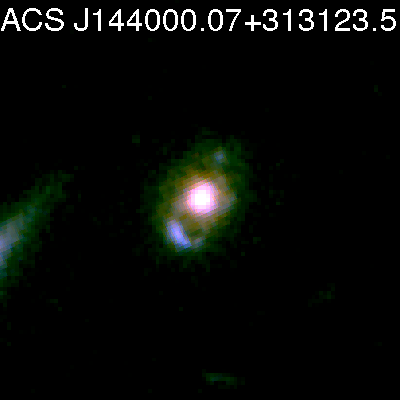}\includegraphics[width= 4.2cm]{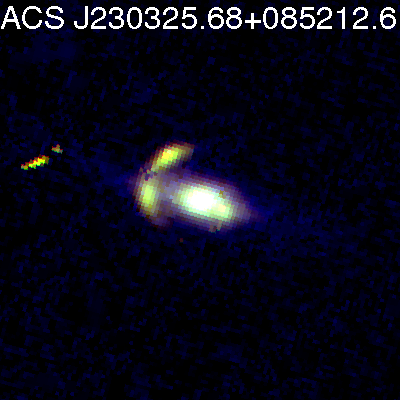}
\caption{Color images of some of our lens candidates. The individual filters used for creating these color images are listed in Table~\ref{tab:colorfilters}. The orientation of the images is north is to the top and East is to the left. The image sizes are similar to  the ones in Figs.~\ref{fig:acscompo1}, \ref{fig:acscompo2} and \ref{fig:acscompo3}  }
\label{fig:ACScolor}
\end{figure*}

A total of 40  new   lens candidates was found in the ACS images  (Figs.~\ref{fig:acscompo1}, \ref{fig:acscompo2} \& \ref{fig:acscompo3}) and nine more in the WFC3 images (Fig.~\ref{fig:WFC3compo1}). Colour  image composites were built for 16 lenses that had observations in at least two filters (Fig.~\ref{fig:ACScolor}), helping to discriminate better between the foreground lens and the source. The filters used to built those colour images are shown in Table~\ref{tab:colorfilters}.  We stress that we  do not  use the colour information as a mean to select the candidates.

\subsection{Lens galaxy surface brightness}\label{sec:galf}

For all our lens candidates, we model and subtract the 2-D light distribution of the lensing galaxy(ies), both to detect any possible counter images for the detected arcs and to measure the magnitude of the lens.

 To  do so, we used  GALFIT v3.0  \citep{2010AJ....139.2097P}  to fit a De Vaucouleur function with elliptical isophotes.  
The code optimizes a number of parameters  describing the lens galaxy: the galaxy total flux,  effective radius (i.e. the half-light radius),   axes ratio and position angle.  Prior to the fit, a Point Spread Function (PSF) was  built using  at least three stars in the vicinity of the lens and using the publicly available PSFEx software \citep{2011ASPC..442..435B}. 
   In order to avoid contamination, the light from the arc(s) and close-by objects was masked out prior to the fit. Images of the  best fit model and of the residuals after subtraction to the original image  are displayed in Figs.~\ref{fig:acscompo1} to \ref{fig:WFC3compo1}. The parameters of the best fit light profiles are given in Tables~\ref{tab:acscan} and \ref{tab:WFC3can}. When needed several galaxies are included in the fit. In this case, several lines are given in the tables.


\subsection{Lens type and classification}\label{class}

Our lens candidates were then classified in different morphological types and a "reliability" criterion is also given to each system. This is summarized in the form of a 3-digit code in the "Comment" column of Tables~\ref{tab:acscan} and \ref{tab:WFC3can}: 
 
\begin{itemize}

\item the first capital letter describes the shape of the source, i.e., whether it is an arc (A), a ring (R) or a compact source (C). 

\item the second character gives number of lensed images seen either in the original image or in the image where the light profile of the lens has been subtracted. 

\item the third, small case letter, is the confidence level for a candidate. This confidence level reflects the author's opinion of how likely a given system is to be a lens or not and is necessarily subjective. The \#a category contains only systems immediately striking to the eye, with image configurations typical for lensing, partial arcs or full Einstein rings. Their morphology alone is strongly suggestive of lensing even without a spectroscopic confirmation. The \#b category is similar to \#a but the morphology alone is not convincing enough. For those, additional spectroscopy, colour images, and lens modeling would help ascertain the lensed nature. Finally, the \#c category contains objects with morphology compatible with lensing but not excluding other interpretations such as chain galaxies, mergers, pairs with tidal tails, or ring galaxies.

\end{itemize}

We find, based solely on morphological criteria that 16 systems fall in the \#a category and 21 systems are in \#b. An additional 12 systems fall in the \#c category. The lens nature of the latter systems must be confirmed with spectroscopy and lens modeling before being fully trusted.  However, if confirmed, they would broaden the variety of known lens morphologies. For this reason, we keep these systems in our catalogue.


\begin{figure}
 \includegraphics[width= 0.5\textwidth]{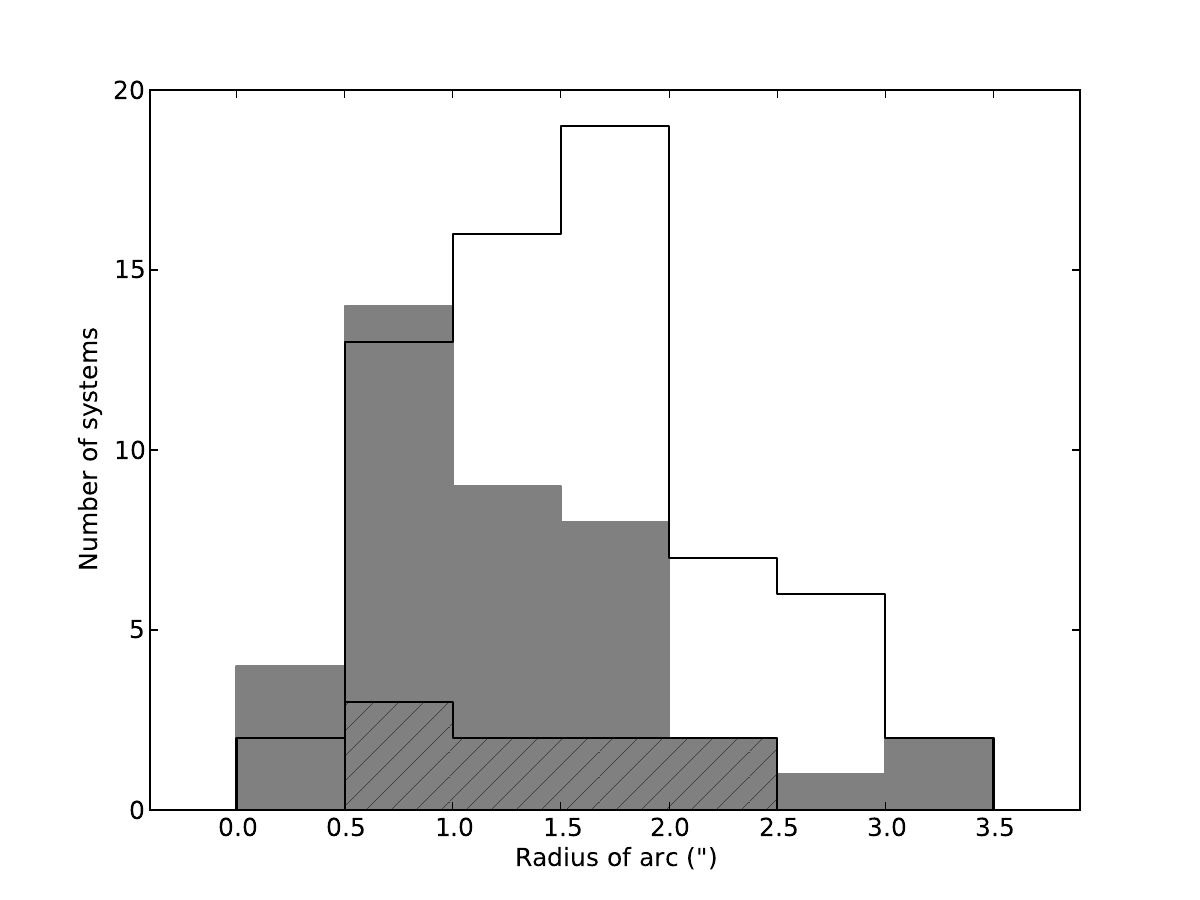}
 \includegraphics[width= 0.5\textwidth]{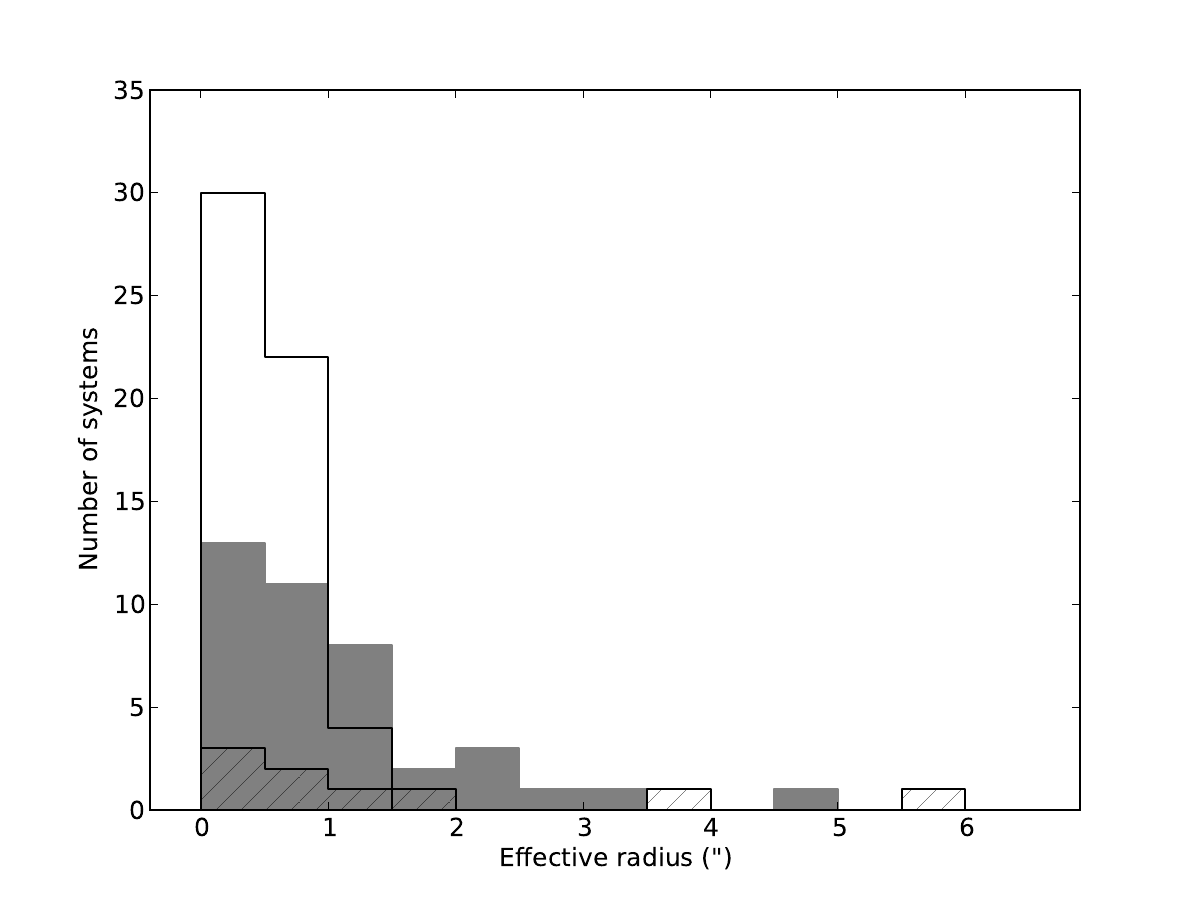}
 \caption{{\it Top:} the arc radii of the ACS lens candidates (filled), WFC3 candidates (hatched) and for the lenses from \citet{Faure2008} for the COSMOS survey (unfilled). {\it Bottom:} comparison of the effective radii of ACS lens candidates (filled), WFC3 candidates (hatched) and of the COSMOS survey (unfilled). For the present sample, when lens candidate is made of two galaxies effective radius of larger lensing galaxy is plotted.}
 \label{fig:radii}
\end{figure}

\section{Comparison with other samples}\label{sec:compa}

\subsection{Comparison to COSMOS}

The new lens search presented in this paper is fully based on the morphology of the candidates in a single band. There is no pre-selection of the best lens candidates in terms of stellar mass or size and we do not use any redshift information. Although  direct comparisons with other lens searches are not straightforward, the work by \citet{Faure2008} in the COSMOS field shares many of our search characteristics, such as the visual inspection of ACS images. 

Indeed the first  lens sample in COSMOS was built by  visually inspecting stamps images of pre-selected bright early-type galaxies. This  affects the relative "success rate" of the two surveys, with COSMOS being biased towards more massive lenses by construction. 
We can nevertheless compare the main morphological characteristics of the two samples of lenses. 
In Fig.~\ref{fig:radii} we compare  the first COSMOS lens sample and  the present sample  arc radius distributions, i.e. the distance between the arc centre of light and the centre of the lens galaxy. In Fig. ~\ref{fig:radii}  we also display  the lens effective radii distributions, using  the best fits in Tables~\ref{tab:acscan} and \ref{tab:WFC3can}. When a lens candidate is made of two galaxies, the effective radius of the largest of the two is considered and the center of the lens is the light barycenter of the two galaxies. This  is the case for only 2 out of the 40 systems, and therefore cannot strongly influence our total lens counts. 


It is immediately seen that the average effective radius of the lens galaxies in our sample is similar to that of COSMOS, but that the average arc radius is smaller in the present sample.  This suggests that for a given effective radius, the lens galaxies in our sample are less massive than in COSMOS. This is not surprising given the pre-selection  of lens candidates done in COSMOS as explained above. 
However, this may also reflect a difference between the distance  $D = D_s / (D_{ls} D_{l})$ between the lenses and the sources of the two samples. We think that this is due to the fixed depth in COSMOS on the one hand, and to the inhomogeneous depth of our dataset on the other hand. Lens modeling would be needed to characterize further the sample. However, in the absence of redshift measurements of our lenses and their respective lensed sources, we do not embark in such a task, which would necessarily be speculative. Given the very different types of lensing objects in the sample we cannot make any reasonable assumption on the lens masses, e.g. using considerations on the fundamental plane. We prefer considering our sample as an empirical way to estimate a lower limit on the number of lenses that may by found in future whole-sky surveys, based purely on morphological criteria. 


From the COSMOS field, \citet{Faure2008, faure2011}  find a  density of $\sim$10 to 20 lenses per square degree down to a depth of $I=26.4$~mag depending on the degree of purity chosen. We find $\sim$7 lenses per square degree  in the present sample. Part of the difference between  COSMOS  and our counts may owe to cosmic variance: COSMOS targets a single line of sight and we use  many random lines of sight (see Fig.~\ref{fig:allsky}). Another reason for the difference is more likely the different requirements on purity between the two samples and the fact that the human time spent per unit area in COSMOS is larger than in the present work. At the depth of COSMOS our survey is already twice larger than the COSMOS field (Fig.~\ref{fig:acsdepth}) and our manpower is limited to a single astronomer for the ACS  6.03\degr$^2$ dataset, and another one for the WFC3 1.01\degr$^2$ dataset. The  search for lenses in the first COSMOS sample was made by four people.

\subsection{Comparison to other surveys}

\begin{figure*}
 \includegraphics[width= 0.45\textwidth]{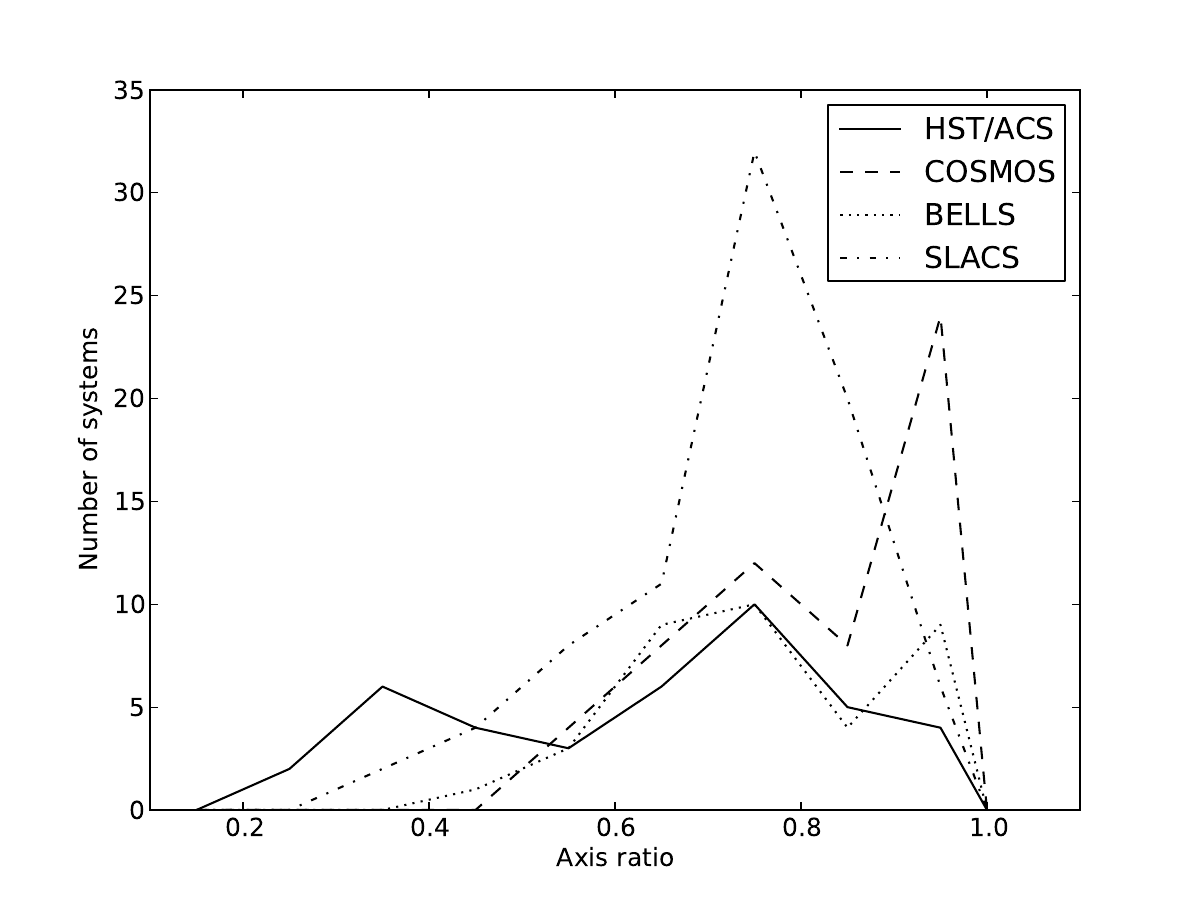}
 \includegraphics[width= 0.45\textwidth]{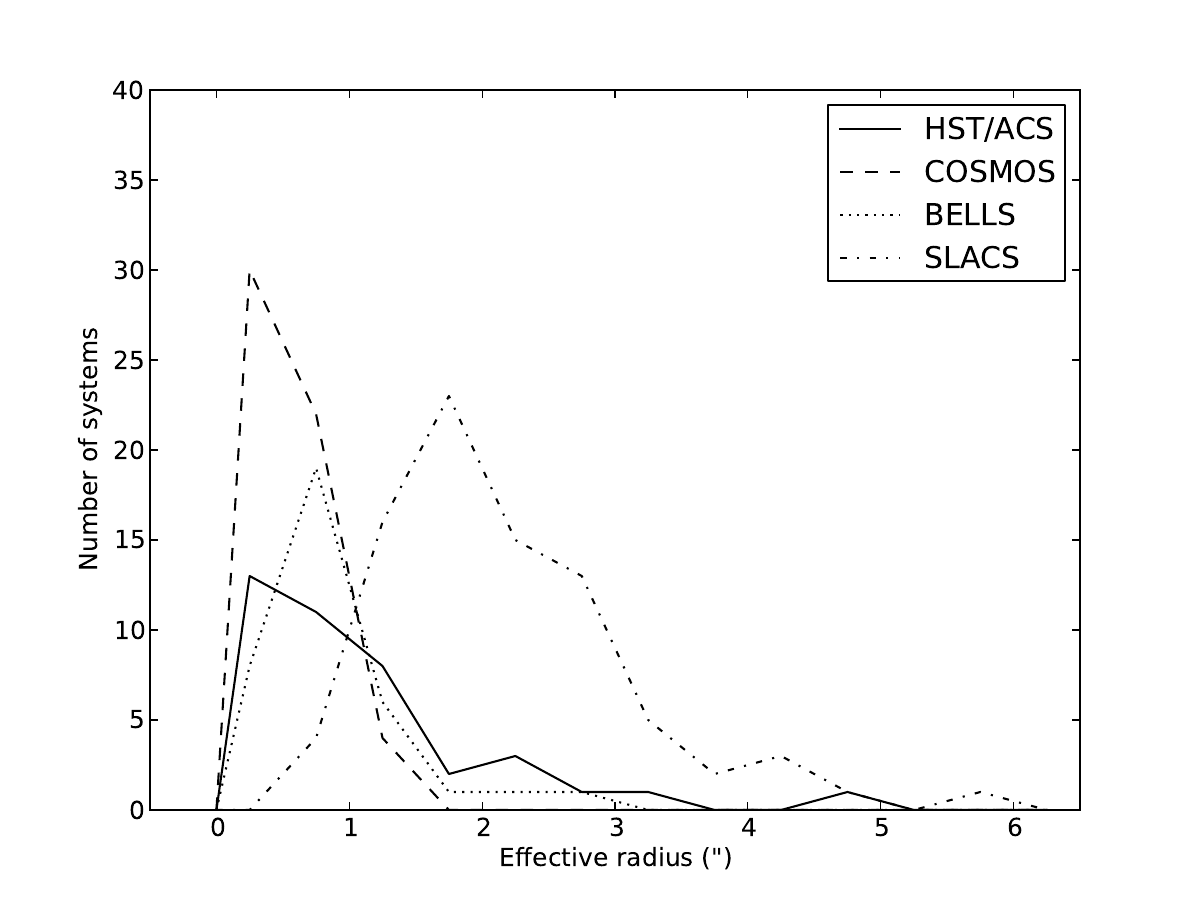}
 \includegraphics[width= 0.45\textwidth]{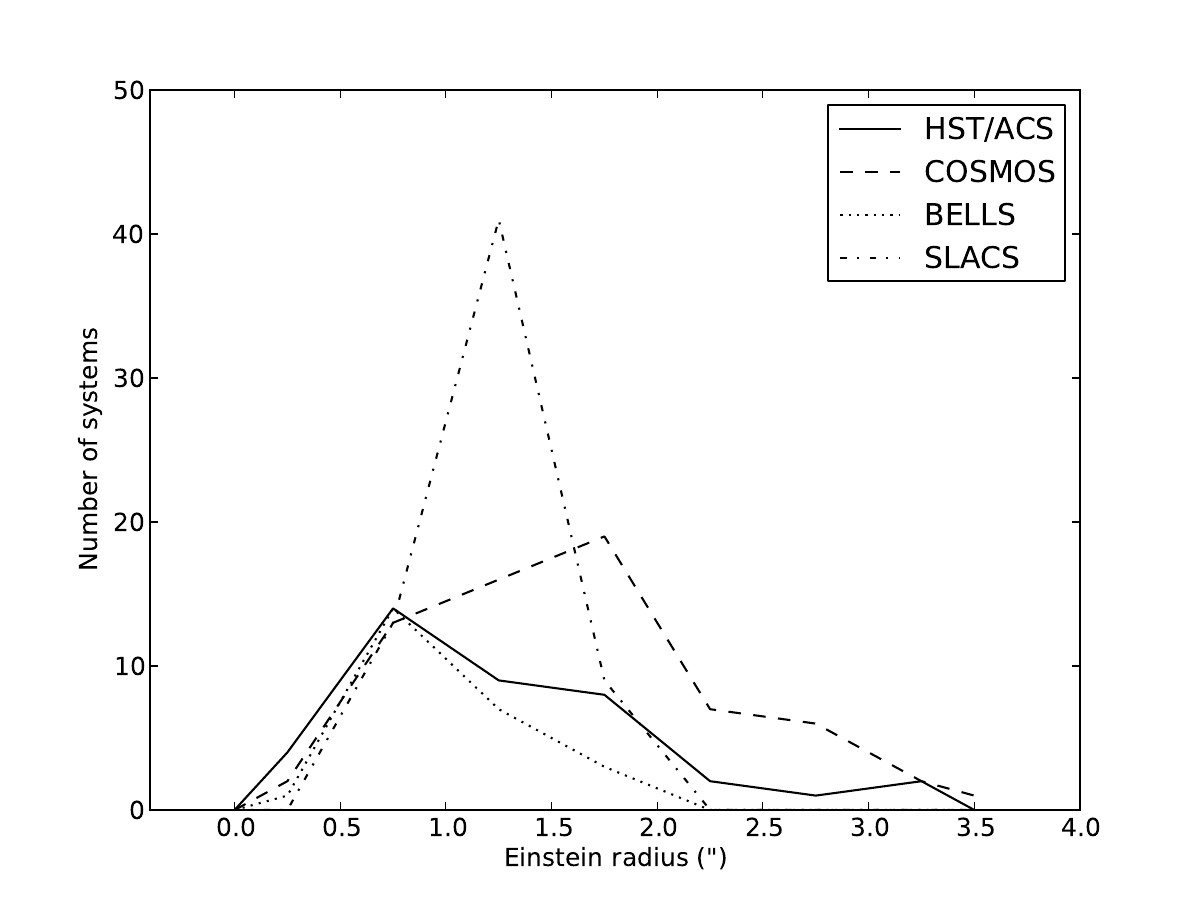}
 \includegraphics[width= 0.45\textwidth]{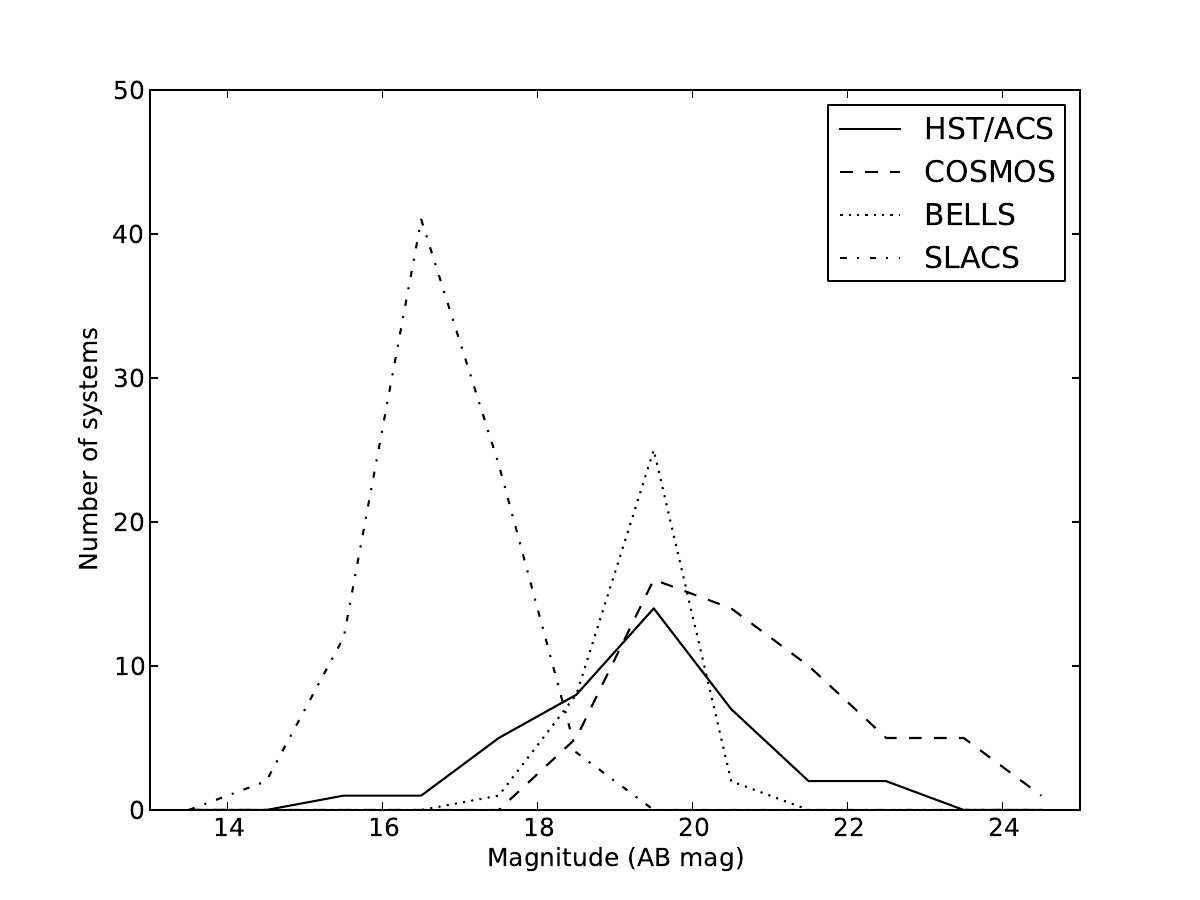}
 \caption{Comparison of the different lens parameters in our study (HST/ACS lenses) and in the COSMOS \citep{Faure2008}, BELLS \citep{2012brownstein} and SLACS \citep{2006ApJ...638..703B, 2008ApJ...682..964B}  lens surveys.}
 \label{fig:compareall}
\end{figure*}

In Fig.~\ref{fig:compareall} we compare the lens properties of some of the largest lens samples available so far: BELLS \citep{2012brownstein}, SLACS \citep{2006ApJ...638..703B, 2008ApJ...682..964B}, COSMOS \citep{Faure2008, Jackson2008}, and the present work. Our new sample has the broadest parameter distributions: axis ratio of the lens, effective radius of the lens, Einstein radius and magnitude of the lens. Although it is hard to infer the exact origin of this effect, it reflects the absence of pre-selection of the lens galaxies in our sample: we inspect every possible object in the HST images. All three other samples involve the pre-selection of the lens galaxies, in mass, morphological type, redshift, magnitudes, sizes. Our sample is therefore more representative of the population of strong lensing systems, which was the original motivation for the visual search. This is to he price, however, of a much more heterogeneous sample and of the impossibility to evaluate the completeness of the sample in an objective way.

We also note that the magnitude distribution of our sample is very similar to the one of BELLS and COSMOS, while SLACS covers a brighter range of magnitudes. Since COSMOS, BELLS and SLACS all are samples of early type massive galaxies this difference can be due to the lower redshift of the SLACS lenses relative to other surveys. This suggests that the lens candidates in our sample have redshifts similar to COSMOS and BELLS. From the radius distributions we see that our sample and BELLS  have similar average effective and Einstein radii, suggesting that they have similar masses and redshifts.

\subsection{Current survey and other serendipitous discoveries}

We have removed the targeted lenses recovered in this survey which were originally discovered in surveys  such as SLACS, as described in section \ref{subsec:inspection}. We also note that one of "our" lenses was serendipitously discovered by \citet{2009ApJ...691..531C}. In addition, a few serendipitous lenses found in data from  HST archive are missed by our survey, like the two ones by \citet{2010ApJS..188..280S}.  One of these lenses is not visually convincing with single band data and the other objects falls by chance on the lines of the grid created to ease inspection of the data. Two more serendipitous lenses from \citet{Fassnacht2006} were missed by our survey. Here as well, the single-band images are not convincing enough. This again highlights the fact that the number counts given in this paper should be considered as lower estimates of the true lens counts.


\begin{table}
   \caption{Number counts of galaxy-scale lenses as a function of depth (a 3-$\sigma$ detection in 1\arcsec$^2$) in the ACS I-band. Note that the magnitudes correspond to the survey depth, not to the magnitude of the faintest lens in a given magnitude bin. The counts are for a 15~000 ${\rm deg}^2$ sky survey (See text for a justification of the Euclid limiting depth).}
   \centering
   \begin{tabular}{c r r}
    \hline
    Depth & Eq.~\ref{eq:all} & Eq.~\ref{eq:dos} \\
    \hline
     24.8 &  34~030          & 19~131  \\
     Euclid (25.8) & 88 756 & 63 629 \\
     26.2 & 128~892         & 101~573  \\
     26.8 & 223~197         & 202~170 \\
\hline
   \end{tabular}
   \label{tab:lenscounts}
\end{table}

\section {Prospects for future space based all-sky surveys}



\begin{figure}
 \includegraphics[width= 0.52\textwidth]{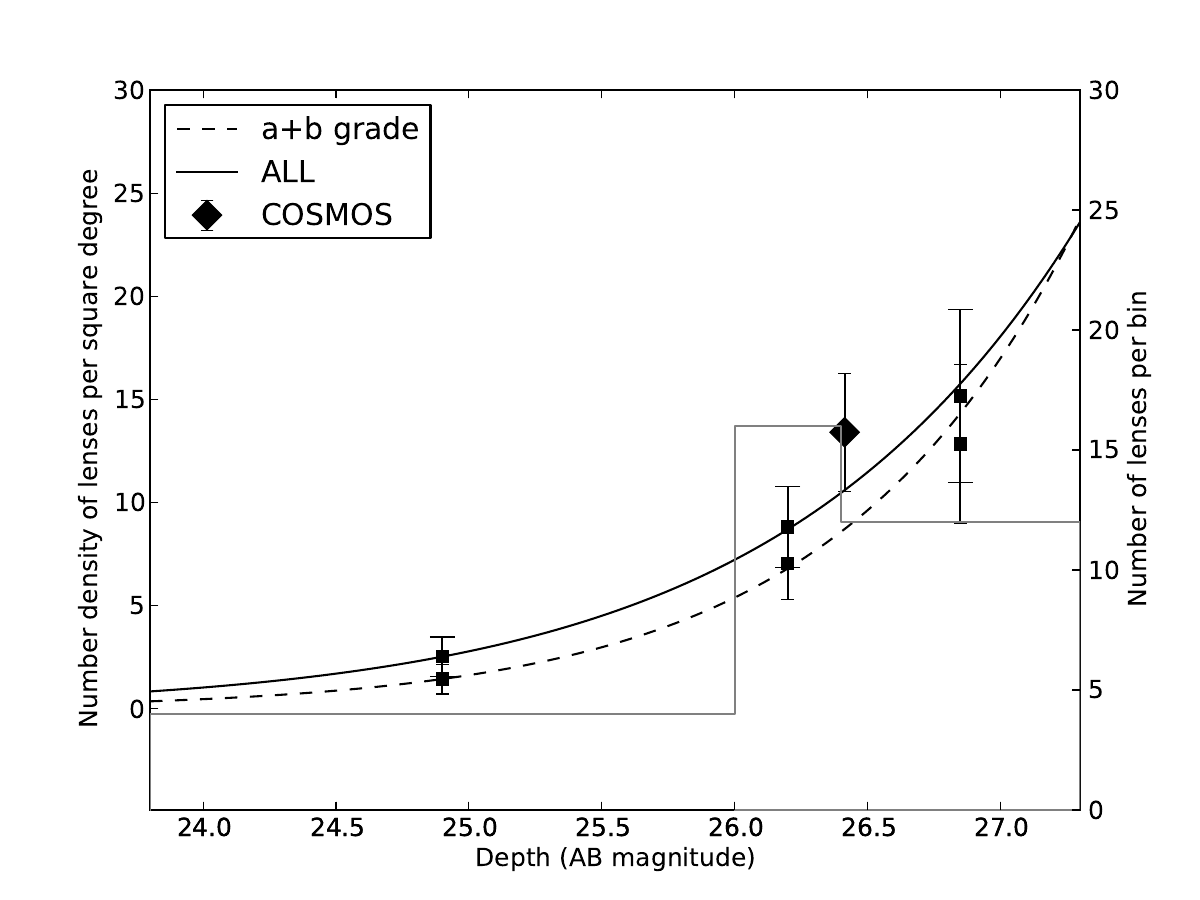}
 \caption{Number density of strong lenses in the HST ACS/F814W archival survey,  where the error bars correspond to the statistical error bars per bin. A total of forty lenses are shown as a function of the survey depth, i.e. each point shows the number of lenses found in a survey of the depth displayed on the {\it x-axis}.  The bins are chosen such that there are enough number of lenses per bin. The histogram in gray shows the bins used and the number of lenses per bin. We plot two curves depending on the confidence level allocated to the lens candidates (see text). For comparison, the diamond-shape point corresponds to the 22 most secure lenses in COSMOS. The limiting depth corresponds to a 3-$\sigma$ detection in 1\arcsec$^2$.}
 \label{fig:lensingrate}
\end{figure}

The specificity of our survey consists in its high spatial resolution, its large field of view, and its variable depth. This allows us to give the lensing rate as a function of depth, as summarized in Fig.~\ref{fig:lensingrate}. In this Figure, we give the number density of strong lenses as a function of the limiting depth. To keep a reasonable number of lenses per magnitude bin, we split our sample in only three bin and we fit a power-law to the data points:


\begin{equation}
 N = A \times m^{\alpha}, 
\end{equation}


which can be used to estimate the expected minimum number density of strong lenses as a function of the survey depth $m$ in AB magnitude in the I-band:

\begin{equation}
N_{all} = 3.47(\pm 11.02)\times10^{-34}\, m^{24.25(\pm4.45)},
\label{eq:all}
\end{equation}

if all our  candidates are included and

\begin{equation}
N_{a+b} =  5.18(\pm 14.32)\times10^{-43}\, m^{30.4(\pm 3.87)},
\label{eq:dos}
\end{equation}

if we keep only the class "a" and class "b" candidates, i.e., the most secure systems. The error bars on the amplitude and index of the power law, given in parentheses, are the result of a least square fit, i.e. they correspond to the internal errors to the fit. We consider this fit as a way to interpolate between our 3 data points rather than any attempt to extrapolate the counts beyond the limiting depth of our survey. In Fig.~\ref{fig:lensingrate}, we also show the COSMOS number density of lenses, where only the 22 most secure systems are considered \citep[see Table~3 in  ][]{faure2011}. The lensing rate is slightly higher  in COSMOS  as discussed in Section~\ref{sec:compa}, but still compatible with ours. 

We can use Eqs.~\ref{eq:all} and \ref{eq:dos} to predict the {\it minimum} lens counts in a 15~000 ${\rm deg}^2$ sky survey. Our estimates are given in Table~\ref{tab:lenscounts}. These counts are for the HST resolution in the I-band. Using these, we can attempt to estimate the number of 
strong lenses expected for a survey like Euclid \citep{Laureijs2011}. To do so, we have degraded our HST images of strong lenses to the Euclid resolution and sampling. This leads to unchanged lens counts compared with the HST, owing to the rather large angular size of all our systems. The limiting magnitude of Euclid will be $m=24.5$ for the optical channel in the AB system. This 10-$\sigma$ limit translates into a 3-$\sigma$ limit of $m=25.8$. According to Eqs.~\ref{eq:all} and \ref{eq:dos}, Euclid should therefore find 88~756 and 63~629 strong lenses, respectively (Table~\ref{tab:lenscounts}). 



It is not possible to provide precise lens counts  per survey depth for our near-IR search, due to the many filters and exposure times applied.

\section{Conclusion}

We have conducted a visual search for galaxy-scale strong gravitational lenses in a total of 7 ${\rm deg}^2$ of the sky, using archival HST images. The data comprise the whole  ACS/WFC F814W imaging data available as of  August 31$^{\rm st}$, 2011 (6.03  ${\rm deg}^2$, excluding the COSMOS field) as well as the WFC3 near-IR images (1.01 ${\rm deg}^2$ in all filters)  up to the same date. 

We have found 49 new lens candidates:  40 in the ACS images and 9 in WFC3 images. Out of these, 16 are  without any doubt genuine lenses, owing to their striking morphology.  An additional 21 are excellent candidates and 12 have morphologies compatible with strong lensing but would need further investigation before being confirmed.
 
Our search for strong lenses is purely based on morphological criteria, with limited colour information for 16 objects out of 49. The colour information is not used to carry out the search itself. Our goal is not to provide a new homogeneous sample well suited for cosmology or galaxy evolution study but is more modest: to estimate a lower limit to the number density of galaxy-scale strong lenses and to illustrate the large diversity in possible image configuration and in possible lens types. Our technique is straightforward and purely based on data rather than on modeling of the population of lens and source galaxies.

The large amount of data in the optical F814W band allows us to estimate the lensing rate as a function of survey depth, at the spatial resolution of the HST. These counts must be taken as lower estimates of the true lens counts owing to the high purity of the sample but also to its incompleteness. Estimating the latter is out of the scope of this paper. Incompleteness can only be estimated with respect to what is believed to be complete, and this requires first to define a precise definition of the lenses to be searched for. In the present case, we look for all possible types of lenses. Defining an incompleteness criterion is therefore an ill-posed problem.

Scaling our lensing rate, we estimate that a 15~000${\degr}^2$ optical survey such as Euclid will find at least 60~000 strong lenses down to I=25.8~mag (3-$\sigma$), which is about  1000 times the current number of known strong lenses in the same area (see the Master lens Database\footnote{http://masterlens.astro.utah.edu/}).


\section{Acknowledgements}
This work is supported by the Swiss National Science Foundation (SNSF). RSP also acknowledge support from a Swiss government scholarship delivered by the Swiss Federal Commission for Scholarships for Foreign Students (CFBE) from June 2010 to June 2012. We also acknowledge support from the International Space Science Institute (ISSI) in Bern, where some of this research has been discussed.

\bibliographystyle{mn2e}
\bibliography{Bibli}
\end{document}